\documentstyle[12pt]{article}

\newcommand {\e} {\mbox{\rm e}}

\newcounter{eq}
\newcounter{sc}

\def\overleftrightarrow#1{\vbox{\ialign{##\crcr
 $\leftrightarrow$\crcr\noalign{\kern-1pt\nointerlineskip}
 $\hfil\displaystyle{#1}\hfil$\crcr}}}

\def\slashb#1{\not\!\!#1}

\newlength{\minitwocolumn}
\begin{document}

\begin{flushright}
DPUR/TH/53\\
October, 2016\\
\end{flushright}
\vspace{20pt}

\pagestyle{empty}
\baselineskip15pt

\begin{center}
{\large\bf Classical Weyl Transverse Gravity
\vskip 1mm }

\vspace{10mm}
Ichiro Oda \footnote{E-mail address:\ ioda@phys.u-ryukyu.ac.jp
}

\vspace{3mm}
           Department of Physics, Faculty of Science, University of the 
           Ryukyus,\\
           Nishihara, Okinawa 903-0213, Japan.\\

\end{center}


\vspace{3mm}
\begin{abstract}
We study various classical aspects of the Weyl transverse (WTDiff) gravity in a general space-time dimension. 
First of all, we clarify a classical equivalence among three kinds of gravitational theories, those are,
the conformally-invariant scalar tensor gravity, Einstein's general relativity and the WTDiff gravity 
via the gauge fixing procedure. Secondly, we show that in the WTDiff gravity the cosmological constant is a mere 
integration constant as in unimodular gravity, but it does not receive any radiative corrections 
unlike the unimodular gravity. A key point in this proof is to construct a covariantly conserved 
energy-momentum tensor, which is achieved on the basis of this equivalence relation. Thirdly, we demonstrate 
that the Noether current for the Weyl transformation is identically vanishing, 
thereby implying that the Weyl symmetry existing in both the conformally-invariant scalar tensor gravity and 
the WTDiff gravity is a "fake" symmetry. We find it possible to extend this proof to all matter fields, i.e. 
the Weyl invariant scalar, vector and spinor fields. 
Fourthly, it is explicitly shown that in the WTDiff gravity the Schwarzshild black hole 
metric and a charged black hole one are classical solutions to the equations of motion 
only when they are expressed in the Cartesian coordinate system.  Finally, we consider 
the Friedmann-Lemaitre-Robertson-Walker (FLRW) cosmology and provide some exact solutions. 
\end{abstract}

\vspace{1mm}

\begin{center}
\it{Dedicated to the memory of Mario Tonin}
\end{center}

\newpage
\pagestyle{plain}
\pagenumbering{arabic}


\rm
\section{Introduction}

The physical importance of Weyl (local conformal) symmetry has not been clearly established in quantum gravity
thus far.  It is usually believed that if the energy scale under consideration goes up to the Planck mass scale, 
all elementary particles, which are either massive or massless at the low energy scale, could be regarded as almost 
massless particles where the Weyl symmetry would become a gauge symmetry and play an important role. 
However, it is true that a concrete implementation of the Weyl symmetry as a plausible gauge symmetry in quantum gravity 
encounters a lot of difficulties. For instance, if one requires an exact Weyl symmetry to be realized in gravitational 
theories at the classical level, only two candidate theories are deserved to be studied though they
possess some defects in their own right. The one theory is the conformal gravity, for which the action is described 
in terms of the square term of the conformal tensor. The conformal gravity belongs to a class of the higher derivative 
gravities so that it suffers from a serious problem, i.e. violation of the unitarity because of the emergence of massive
ghosts although it has an attractive feature as a renormalizable theory \cite{Stelle, Julve}. 

The other plausible candidate as a gravitational theory with the Weyl symmetry, which we consider in this article 
intensively, is the conformally invariant scalar-tensor gravity \cite{Dirac, Deser}. In this theory, 
a (ghost-like) scalar field is introduced in such a way that it couples to the scalar curvature in a conformally invariant 
manner. Even if this theory is a unitary theory owing to the presence of only second-order derivative terms, 
it suffers from a sort of triviality problem in the sense that 
when we take a suitable gauge condition for the Weyl symmetry (we take the scalar field to be a costant), the action of 
the conformally invariant scalar-tensor gravity reduces to the Einstein-Hilbert action of Einstein's general relativity. 
It is therefore unclear to make use of the conformally invariant scalar-tensor gravity as an alternative theory 
of general relativity. Of course, the conformally invariant scalar-tensor gravity is not a renormalizable theory like
general relativity.

One reason why we would like to consider a gravitational theory with the Weyl symmetry stems from the cosmological
constant problem \cite{Weinberg}, which is one of the most difficult problems in modern theoretical physics. The Weyl symmetry 
forbids the appearance of operators of dimension zero such as the cosmological constant in the action so it is expected that 
the Weyl symmetry might play an important role in the cosmological constant problem \cite{Oda0}. In this respect, a difficulty 
is that the Weyl symmetry is broken by quantum effects and its violation emerges as a trace anomaly of the energy-momentum
tensor \cite{Capper1, Capper2}. Thus, the idea such that one utilizes the Weyl symmetry as a resolution of the cosmological 
constant problem makes no sense at the quantum level even if it is an intriguing idea at the classical level.

Here a naive but natural question arises: Is the Weyl symmetry always violated by radiative corrections? We think that
it is not always so. What kind of the Weyl symmetry is not broken? In a pioneering work by Englert's et al. \cite{Englert}, 
it has been clarified that the conformally invariant scalar-tensor gravity coupled to various matter fields is free of
Weyl anomaly when the Weyl symmetry is spontaneously broken. This fact has been investigated and certified by subsequent 
papers \cite{Shaposhnikov1}-\cite{Ghilencea2}. Related to these works, in this article, we wish to put forward 
a new conjecture that the Weyl symmetry is not violated by radiative corrections if it is a ${\it fake}$ Weyl symmetry
and it is spontaneously broken. Here the word "${\it fake}$" means that the corresponding Noether currents \cite{Noether} 
vanish identically \cite{Jackiw, Oda0}.

If our conjecture were really valid, we could have recourse to the Weyl symmetry as a resolution to the cosmological
constant problem as follows: Start with the conformally invariant scalar-tensor gravity, and gauge-fix the longitudinal 
diffeomorphism instead of the Weyl symmetry, by which the ghost-like scalar field can be removed from the physical spectrum, 
so that the unitarity issue does not occur. Consequently, we obtain the Weyl invariant and transverse diffeomorphisms-invariant 
gravitational theory. We then find that the remaining Weyl symmetry is a fake symmetry so it is not violated by quantum 
corrections according to our conjecture. By a detailed analysis, it turns out that this gravitational theory, 
which we call Weyl transverse (WTDiff) gravity \cite{Izawa}-\cite{Oda3}, has a remarkable feature that the equations of motion 
can be rewritten to the same form as the standard Einstein's equations where the cosmological constant emerges 
as an integration constant as in unimodular gravity. In the unimodular gravity \cite{Einstein}-\cite{Nojiri2}, 
the unimodular condition is implemented by using the Lagrange multiplier field, 
which plays a role as the cosmological constant and receives huge radiative corrections, so the cosmological constant poblem 
is not solved. On the other hand, in the WTDiff gravity, there is no constraint like the unimodular condition and the
unbroken Weyl symmetry severely prohibits the appearance of the cosmological constant. Hence, our conjecture would insist
that in the WTDiff gravity, the cosmological constant problem is reduced to a mere problem of how to fix the initial value 
of the cosmological constant, which is an important first step for a resolution of the cosmological constant problem
though we still have a new problem of how to fix its initial vaule.  
  
This paper is organised as follows: In Section 2, we clarify the equivalence relation among three kinds of gravitational 
theories, i.e. the conformally-invariant scalar tensor gravity, Einstein's general relativity and the WTDiff gravity 
via the gauge fixing procedure. This equivalence makes it possible to construct a covariantly conserved energy-momentum
tensor and prove that the equations of motions in the WTDiff gravity can be transformed to the Einstein equations of
general relativity. The possibility of making such an energy-momentum tensor comes from the fact that the underlying
theory behind the WTDiff gravity is the conformally invariant scalar-tensor gravity which is generally covariant.

In Section 3, we show that the Noether current for the Weyl transformation is identically 
vanishing, thereby implying that the Weyl symmetry existing in both the conformally-invariant scalar tensor gravity and 
the WTDiff gravity is a "fake" symmetry. It is shown that it is possible to apply this proof for 
all the Weyl invariant matter fields. 
It is explicitly shown in Sections 4 and 5 that in the WTDiff gravity the Schwarzschild black hole 
metric and the charged black hole one are classical solutions to the equations of motion 
only when they are expressed in the Cartesian coordinate system.  In Section 6, we consider 
the Friedmann-Lemaitre-Robertson-Walker (FLRW) cosmology and provide an exact solution. 
The final section is devoted to discussions. Our notation and conventions are summarized in Appendix A. From Appendix B
to D, some proof and the details of calculations are presented.

\section{Equivalence among three gravitational theories}

We will start by recalling the well-known recipe for obtaining the conformally invariant scalar-tensor
gravity from the Einstein-Hilbert action of general relativity. The Einstein-Hilbert action is of form
in a general $n$ space-time dimension (We assume $n \not= 2$ in this article.) \footnote{See Appendix A.1
for our notation and conventions.} 
\begin{eqnarray}
\hat S = \frac{1}{2} \int d^n x \sqrt{- \hat g} \hat R,
\label{E-H Action}
\end{eqnarray}
where $\hat g_{\mu\nu}$ is a metric tensor. (The "hat" symbol is put for later convenience.) 
To let this action have the Weyl (local conformal) symmetry, one introduces a scalar field $\varphi$ 
and supposes that the metric tensor $\hat g_{\mu\nu}$ is composed of the scalar field
$\varphi$ and a new metric field $g_{\mu\nu}$ as
\begin{eqnarray}
\hat g_{\mu\nu} = \left( \frac{1}{2} \sqrt{\frac{n-2}{n-1}} \varphi \right)^{\frac{4}{n-2}} g_{\mu\nu}.
\label{Weyl invariant metric}
\end{eqnarray}
The key observation is that the metric tensor $\hat g_{\mu\nu}$ is invariant under the following Weyl transformation
\begin{eqnarray}
g_{\mu\nu} \rightarrow g^\prime_{\mu\nu} = \Omega^2(x) g_{\mu\nu}, \quad
\varphi \rightarrow \varphi^\prime = \Omega^{-\frac{n-2}{2}}(x) \varphi,
\label{Weyl transf}
\end{eqnarray}
where $\Omega(x)$ is a scalar parameter. Next, substituting (\ref{Weyl invariant metric}) into 
the Einstein-Hilbert action (\ref{E-H Action}) produces an action for the conformally invariant scalar-tensor gravity
\footnote{See Ref. \cite{Oda7}-\cite{Oda10} for various applications of this action.}
\begin{eqnarray}
S = \int d^n x \ \sqrt{-g} \left[ \frac{n-2}{8(n-1)} \varphi^2 R +  \frac{1}{2}
g^{\mu\nu} \partial_\mu \varphi \partial_\nu \varphi  \right].
\label{Cof-inv S-T Action}
\end{eqnarray}
Note that the scalar field $\varphi$ is not normal but ghost-like owing to the positive coefficient $\frac{1}{2}$,
but it is not a problem since the dynamical degree of freedom associated with $\varphi$ can be nullified by taking
a gauge condition.

This recipe for introducing the Weyl symmetry to a theory suggests that the Weyl symmetry obtained in this way 
might be a {\it {fake}} symmetry and the scalar field $\varphi$ be a spurion field \cite{Jackiw}. Indeed, as shown later,
the Noether current for the Weyl symmetry is identically vanishing for both local and global Weyl transformations 
\cite{Jackiw, Oda0}.
The physical property and the importance of this {\it {fake}} Weyl symmetry will be also discussed later in
dealing with the Noether currents.     

Now let us invert the order of the above argument and this time start with the conformally invariant scalar-tensor
gravity (\ref{Cof-inv S-T Action}). It is easy to see that a gauge condition for the Weyl symmetry
\begin{eqnarray}
\varphi = 2 \sqrt{\frac{n-1}{n-2}},
\label{GR gauge}
\end{eqnarray}
transforms the action (\ref{Cof-inv S-T Action}) of the conformally invariant scalar-tensor gravity into
the Einstein-Hilbert action (\ref{E-H Action}) of general relativity (without the hat symbol).   
 
An interesting gauge condition for not the Weyl symmetry but the longitudinal diffeomorphism is given by
\begin{eqnarray}
\varphi = 2 \sqrt{\frac{n-1}{n-2}} |g|^{- \frac{n-2}{4n}},
\label{WTDiff gauge}
\end{eqnarray}
where we have defined $|g| = -g$ because of $g < 0$. Here let us examine this gauge condition (\ref{WTDiff gauge})
more closely. Under the Weyl transformation (\ref{Weyl transf}), the RHS of Eq. (\ref{WTDiff gauge}) is transformed
as
\begin{eqnarray}
2 \sqrt{\frac{n-1}{n-2}} |g|^{- \frac{n-2}{4n}} \rightarrow \Omega^{- \frac{n-2}{2}} \ 2 \sqrt{\frac{n-1}{n-2}} 
|g|^{- \frac{n-2}{4n}},
\label{Weyl transf of WTDiff gauge}
\end{eqnarray}
which is the same transformation property as $\varphi$ under the Weyl transformation as seen in (\ref{Weyl transf}).
Thus, the gauge condition (\ref{WTDiff gauge}) does not break the Weyl symmetry. Instead, the gauge condition 
(\ref{WTDiff gauge}) {\it{does}} break the longitudinal diffeomorphism as explained in what follows: 
First, notice that with the gauge condition (\ref{WTDiff gauge}) the metric tensor (\ref{Weyl invariant metric})
reads
\begin{eqnarray}
\hat g_{\mu\nu} = |g|^{-\frac{1}{n}} g_{\mu\nu}.
\label{WTDiff gauge metric}
\end{eqnarray}
Taking the determinant of this metric reveals us that $\hat g_{\mu\nu}$ is the unimodular metric satisfying
the unimodular condition 
\begin{eqnarray}
\hat g(x) = -1.
\label{Unimodularity}
\end{eqnarray}
Given the unimodular condition (\ref{Unimodularity}), any variation of the unimodular metric gives rise to
an equation  
\begin{eqnarray}
\hat g^{\mu\nu} \delta \hat g_{\mu\nu} = 0.
\label{Traceless variation}
\end{eqnarray}
When one restricts the variation to be diffeomorphisms
\begin{eqnarray}
\delta \hat g_{\mu\nu} = \hat \nabla_\mu \xi_\nu + \hat \nabla_\nu \xi_\mu
=  \hat g_{\mu\rho} \partial_\nu \xi^\rho + \hat g_{\nu\rho} \partial_\mu \xi^\rho
+ \xi^\rho \partial_\rho \hat g_{\mu\nu},
\label{GCT}
\end{eqnarray}
with $\hat \nabla_\mu$ and $\xi_\mu$ being the covariant derivative with respect to the metric tensor
$\hat g_{\mu\nu}$ and an infinitesimal parameter, respectively, Eq. (\ref{Traceless variation})
yields 
\begin{eqnarray}
\partial_\mu \xi^\mu = 0,
\label{TDiff}
\end{eqnarray}
where we have used 
\begin{eqnarray}
\hat g^{\mu\nu} \partial_\rho \hat g_{\mu\nu} = 2 \partial_\rho \left( \log \sqrt{- \hat g} \right)  = 0,
\label{Formula from Unimod}
\end{eqnarray}
which comes from the unimodular condition (\ref{Unimodularity}). The equation (\ref{TDiff}) implies that 
the full group of diffeomorphisms (Diff) is broken down to the transverse diffeomorphisms (TDiff) \footnote{See
Appendix B for more details of TDiff.}, thereby showing that the gauge condition (\ref{WTDiff gauge}) certainly breaks 
the longitudinal diffeomorphism. 

Inserting the gauge condition (\ref{WTDiff gauge}) to the action of the conformally invariant scalar-tensor gravity
(\ref{Cof-inv S-T Action}), one arrives at an action of the Weyl transverse (WTDiff) gravity 
\begin{eqnarray}
S &=& \int d^n x {\cal{L}}     \nonumber\\
&=& \frac{1}{2} \int d^n x |g|^{\frac{1}{n}} \left[ R + \frac{(n-1)(n-2)}{4n^2} \frac{1}{|g|^2}
g^{\mu\nu} \partial_\mu |g| \partial_\nu |g|  \right].
\label{WTDiff Action}
\end{eqnarray}
It is straightforward to derive the equations of motion from this action. The detailed calculation is presented
in the Appendix C by means of two different methods. Then, the equations of motion read
\begin{eqnarray}
R_{\mu\nu} - \frac{1}{n} g_{\mu\nu} R = T_{(g) \mu\nu} - \frac{1}{n} g_{\mu\nu} T_{(g)},
\label{Eq from WTDiff Action}
\end{eqnarray}
where the energy-momentum tensor $T_{(g) \mu\nu}$ is defined as
\begin{eqnarray}
T_{(g) \mu\nu} = \frac{(n-2)(2n-1)}{4n^2} \frac{1}{|g|^2} \partial_\mu |g| \partial_\nu |g|
- \frac{n-2}{2n} \frac{1}{|g|} \nabla_\mu \nabla_\nu |g|,
\label{T(g)}
\end{eqnarray}
with being defined as $\nabla_\mu \nabla_\nu |g| = \partial_\mu \partial_\nu |g| - \Gamma^\rho_{\mu\nu} 
\partial_\rho |g|$. Note that Eq. (\ref{Eq from WTDiff Action}) is purely the traceless part of the
standard Einstein equations. By an explicit calculation, it is possible to verify that the action (\ref{WTDiff Action})
and the equations of motion (\ref{Eq from WTDiff Action}) are invariant under the Weyl transformation (\ref{Weyl transf})
and the transverse group of diffeomorphisms. The proof is given in Appendix B. 

The most important point associated with this energy-momentum tensor (\ref{T(g)}) is that 
it is not covariantly conserved  
\begin{eqnarray}
\nabla^\mu T_{(g) \mu\nu} \not= 0.
\label{Non-conserv T(g)}
\end{eqnarray}
This is because the WTDiff gravity action (\ref{WTDiff Action}) is not invariant under the full group of
diffeomorphisms but only its subgroup, that is, the transverse diffeomorphisms (TDiff). However, our starting 
action (\ref{Cof-inv S-T Action}) of the conformally invariant scalar-tensor gravity is generally covariant, 
so it it should be possible to find an alternative energy-momentum tensor which is covariantly conserved. 
(See Appendix D.)

To find the desired energy-momentum tensor, let us begin by deriving the equations of motion of the
conformally invariant scalar-tensor gravity (\ref{Cof-inv S-T Action}). The equations of motion for the
metric tensor $g_{\mu\nu}$ and the scalar field $\varphi$ are respectively given by
\begin{eqnarray}
\frac{n-2}{8(n-1)} \left[ \varphi^2 G_{\mu\nu} + ( g_{\mu\nu} \Box - \nabla_\mu \nabla_\nu ) 
(\varphi^2)\right] = \frac{1}{4} g_{\mu\nu} \partial_\rho \varphi \partial^\rho \varphi 
- \frac{1}{2} \partial_\mu \varphi \partial_\nu \varphi,
\label{g-Eq of motion of conf-ST}
\end{eqnarray}
and 
\begin{eqnarray}
\frac{n-2}{4(n-1)} \varphi R = \Box \varphi,
\label{varphi-Eq of motion of conf-ST}
\end{eqnarray}
where $G_{\mu\nu} = R_{\mu\nu} - \frac{1}{2} g_{\mu\nu} R$ is the Einstein tensor 
and $\Box \varphi = g^{\mu\nu} \nabla_\mu \nabla_\nu \varphi$. Eq. (\ref{varphi-Eq of motion of conf-ST})
is the equation of motion for the spurion field $\varphi$ so it is not an independent equation. Actually,
taking the trace part of Eq. (\ref{g-Eq of motion of conf-ST}) naturally leads to Eq. (\ref{varphi-Eq of motion of conf-ST}).
Thus, it is sufficient to take only the equations of motion (\ref{g-Eq of motion of conf-ST}) into consideration.

Next, we will rewrite (\ref{g-Eq of motion of conf-ST}) as
\begin{eqnarray}
G_{\mu\nu} &=& \frac{1}{\varphi^2} ( \nabla_\mu \nabla_\nu - g_{\mu\nu} \Box ) (\varphi^2) 
+ \frac{8(n-1)}{n-2} \frac{1}{\varphi^2} \left[ \frac{1}{4} g_{\mu\nu} \partial_\rho \varphi \partial^\rho \varphi 
- \frac{1}{2} \partial_\mu \varphi \partial_\nu \varphi \right]     \nonumber\\
&=& T_{\mu\nu},
\label{g-Eq of motion of conf-ST 2}
\end{eqnarray}
where we have defined a new energy-momentum tensor $T_{\mu\nu}$. Since the Einstein tensor $G_{\mu\nu}$ satisfies
the Bianchi identity
\begin{eqnarray}
\nabla^\mu G_{\mu\nu} = 0,
\label{Bianchi}
\end{eqnarray}
the new energy-momentum $T_{\mu\nu}$ should satisfy the covariant conservation law 
\begin{eqnarray}
\nabla^\mu T_{\mu\nu} = 0.
\label{Conserved T}
\end{eqnarray}

Finally, substituting the gauge condition (\ref{WTDiff gauge}) into $T_{\mu\nu}$, one has
\begin{eqnarray}
T_{\mu\nu} = T_{(g)\mu\nu} + \frac{n-2}{2n} g_{\mu\nu} \left[ - \frac{5n-3}{4n} \frac{1}{|g|^2} 
(\partial_\rho |g|)^2 + \frac{1}{|g|} \nabla_\rho \nabla^\rho |g| \right].
\label{T}
\end{eqnarray}
Note that the existence of the extra terms except $T_{(g)\mu\nu}$ makes it possible to hold the covariant
conservation law (\ref{Conserved T}). Indeed, it is straightforward to check that this energy-momentum
tensor (\ref{T}) satisfies the covariant conservation law (\ref{Conserved T}) by a direct calculation.
Another indirect but easy proof is to consider the conservation law (\ref{Conserved T}) in the local Lorentz frame 
where $g_{\mu\nu} = \eta_{\mu\nu}$ and $\partial_\rho g_{\mu\nu} = 0$. Then, we can explicitly check that
\begin{eqnarray}
\partial^\mu T_{\mu\nu} = - \frac{n-2}{2n} \frac{1}{|g|} \partial^\mu \partial_\mu \partial_\nu |g|
+ \frac{n-2}{2n} \frac{1}{|g|} \partial^\mu \partial_\mu \partial_\nu |g| = 0,
\label{Conserved T in LLF}
\end{eqnarray}
which implies the conservation law (\ref{Conserved T}) in a curved space-time.

One remarkable feature of this energy-momentum tensor $T_{\mu\nu}$
is that there exists a nontrivial relation between $T_{\mu\nu}$ and $T_{(g)\mu\nu}$
\begin{eqnarray}
T_{\mu\nu} - \frac{1}{n} g_{\mu\nu} T = T_{(g)\mu\nu} - \frac{1}{n} g_{\mu\nu} T_{(g)},
\label{Critical relation of T}
\end{eqnarray}
which stems from the fact that the actions of both the conformally invariant scalar-tensor gravity and the WTDiff
gravity are invariant under the Weyl transformation. 
It is worthwhile to stress that our findings (\ref{T}) critically depend on the classical equivalence between the conformally 
invariant scalar-tensor gravity and the WTDiff gravity. In other words, without this equivalence, it would be difficult, 
if not impossible, to construct the covariantly conserved energy-momentum tensor (\ref{T}) which also satisfies the important 
relation (\ref{Critical relation of T}).

Now we are ready to show that the equations of motion of the WTDiff gravity, Eq. (\ref{Eq from WTDiff Action}),
reproduce the standard Einstein equations. To this aim, using Eq. (\ref{Critical relation of T}), let us first replace 
the RHS in Eq. (\ref{Eq from WTDiff Action}) with its covariantly conserved counterpart 
\begin{eqnarray}
R_{\mu\nu} - \frac{1}{n} g_{\mu\nu} R = T_{\mu\nu} - \frac{1}{n} g_{\mu\nu} T.
\label{Cov-Eq from WTDiff Action}
\end{eqnarray}
Taking the covariant derivative of this equation, and using the Bianchi identity (\ref{Bianchi}) and the covariant
conservation law (\ref{Conserved T}), one obtains
\begin{eqnarray}
\frac{n-2}{2n} \nabla_\mu R = - \frac{1}{n} \nabla_\mu T. 
\label{Einstein step 1}
\end{eqnarray}
This equation says that $R + \frac{2}{n-2} T$ is a constant, which we will call $\frac{2n}{n-2} \Lambda$,
\begin{eqnarray}
R + \frac{2}{n-2} T = \frac{2n}{n-2} \Lambda. 
\label{Einstein step 2}
\end{eqnarray}
Eliminating $T$ from Eq. (\ref{Cov-Eq from WTDiff Action}) in terms of Eq. (\ref{Einstein step 2}), one can
reach the standard Einstein equations 
\begin{eqnarray}
R_{\mu\nu} - \frac{1}{2} g_{\mu\nu} R + \Lambda g_{\mu\nu} = T_{\mu\nu}.
\label{Einstein eq}
\end{eqnarray}

Although we have obtained the Einstein equations from the equations of motion of the WTDiff gravity in 
this way, the cosmological constant $\Lambda$ emerges as a mere integration constant and has nothing to
do with any terms in the action or vacuum fluctuations. To put differently, Eq. (\ref{Cov-Eq from WTDiff Action})
does not include the cosmological constant and the contribution from radiative corrections to the cosmological
constant cancels in the RHS of Eq. (\ref{Cov-Eq from WTDiff Action}), thereby guaranteeing the stability of the
cosmological constant against quantum corrections. 

This feature of the emergence of the cosmological constant as an integration constant is a common feature of the WTDiff
gravity and the unimodular gravity \cite{Einstein}-\cite{Nojiri2}. However, there is an important difference between 
the two theories. In the unimodular gravity, from the viewpoint of quantum field
theories, the unimodular condition must be properly implemented via the Lagrange multiplier field, which turns out
to correspond to the cosmological constant in the unimodular gravity, thereby rendering its initial value 
radiatively unstable. In this sense, the cosmological constant problem cannot be solved within the framework of
the unimodular gravity. 

On the other hand, in the WTDiff gravity, there is no constraint like the unimodular condition
and the {\it{fake}} Weyl symmetry is expected to forbid operators of dimension zero such
as the cosmological constant. Moreover, we have a plausible conjecture such that the {\it{fake}} Weyl symmetry
might never be violated by quantum effects, that is, no Weyl anomaly, owing to its "fakeness". In other words,
the {\it{fake}} Weyl symmetry could survive even at the quantum level, by which suppressing the radiative 
corrections to the cosmological constant. If our conjecture were true, the cosmological constant problem
would amount to be a mere problem of how to fix the integration constant $\Lambda$. From this point of view,
we should clarify quantum aspects of the WTDiff gravity in future.

\section{Fake Weyl symmetry}

In our previous article \cite{Oda0}, motivated with the article \cite{Jackiw} we have studied Weyl symmetry (local conformal symmetry) 
in the WTDiff gravity and the conformally invariant scalar-tensor gravity in four space-time dimensions, and shown that 
the Noether currents for both the local and global Weyl symmetries are identically vanishing. In this sense, the Weyl symmetry
existing in the WTDiff gravity and the conformally invariant scalar-tensor gravity is called a "fake" Weyl symmetry. 
In this section, we will generalize this study to not only an arbitrary space-time dimension but also all matter fields
involving scalar, vector and spinor fields.      

\subsection{Gravity}

Let us start with a gravitational sector and consider the action of the WTDiff gravity, Eq. (\ref{WTDiff Action}).
Here it is more convenient to work with the Langrangian density than the action itself
\begin{eqnarray}
{\cal L} = \frac{1}{2} |g|^{\frac{1}{n}} \left[ R + \frac{(n-1)(n-2)}{4n^2} \frac{1}{|g|^2}
g^{\mu\nu} \partial_\mu |g| \partial_\nu |g|  \right],
\label{WTDiff Lagr}
\end{eqnarray}
which is invariant under the Weyl transformation up to a surface term as shown in Appendix B.

Now we wish to calculate the Noether current for Weyl symmetry by using the Noether procedure \cite{Noether}.
We will closely follow the line of arguments in Ref. \cite{Jackiw}. 
The general variation of the Lagrangian density (\ref{WTDiff Lagr}) reads
\begin{eqnarray}
\delta {\cal L} =  \frac{\partial {\cal L}}{\partial g_{\mu\nu}} \delta g_{\mu\nu}
+ \frac{\partial {\cal L}}{\partial ( \partial_\mu g_{\nu\rho} )} \delta (\partial_\mu g_{\nu\rho})
+ \frac{\partial {\cal L}}{\partial ( \partial_\mu \partial_\nu g_{\rho\sigma} )} 
\delta ( \partial_\mu \partial_\nu g_{\rho\sigma} ).
\label{Var-L}
\end{eqnarray}
In this expression, let us note that the Lagrangian density under consideration includes second-order derivatives 
of $g_{\mu\nu}$ in the scalar curvature $R$. Setting $\Omega(x) = \e^{- \Lambda(x)}$, the infinitesimal variation 
$\delta {\cal L}$ under the Weyl transformation (\ref{Weyl transf}) is given by
\begin{eqnarray}
\delta {\cal L} =  \partial_\mu X_1^\mu,
\label{delta L1}
\end{eqnarray}
where $X_1^\mu$ is defined as 
\begin{eqnarray}
X_1^\mu = (n-1) |g|^{\frac{1}{n}} g^{\mu\nu} \partial_\nu \Lambda.
\label{X1}
\end{eqnarray}

Next, using the equations of motion
\begin{eqnarray}
\frac{\partial {\cal L}}{\partial g_{\mu\nu}} 
= \partial_\rho \frac{\partial {\cal L}}{\partial ( \partial_\rho g_{\mu\nu} )}
- \partial_\rho \partial_\sigma \frac{\partial {\cal L}}{\partial ( \partial_\rho \partial_\sigma g_{\mu\nu} )}, 
\label{Eq.M}
\end{eqnarray}
the variation $\delta {\cal L}$ in (\ref{Var-L}) can be cast to the form
\begin{eqnarray}
\delta {\cal L} =  \partial_\mu K_1^\mu,
\label{delta L2}
\end{eqnarray}
where $K_1^\mu$ is defined as 
\begin{eqnarray}
K_1^\mu = \frac{\partial {\cal L}}{\partial ( \partial_\mu g_{\nu\rho} )} \delta g_{\nu\rho}
+ \frac{\partial {\cal L}}{\partial ( \partial_\mu \partial_\nu g_{\rho\sigma} )} \partial_\nu \delta g_{\rho\sigma}
- \partial_\nu \frac{\partial {\cal L}}{\partial ( \partial_\mu \partial_\nu g_{\rho\sigma} )} \delta g_{\rho\sigma}.
\label{K1}
\end{eqnarray}
Using this formula, an explicit calculation yields
\begin{eqnarray}
K_1^\mu = X_1^\mu,
\label{K1=X1}
\end{eqnarray}
thereby giving us the result that the Noether current for the Weyl symmetry vanishes identically 
\begin{eqnarray}
J_1^\mu = K_1^\mu - X_1^\mu = 0.
\label{J1}
\end{eqnarray}
Incidentally, let us note that both the expressions $X_1^\mu$ and $K_1^\mu$ are gauge invariant under 
the Weyl transformation as seen in Eq. (\ref{X1}). This fact will be utilized later.

As an alternative derivation of the same result, one can also appeal to a more conventional
method where the Lagrangian density in (\ref{WTDiff Lagr}) does not explicitly
involve second-order derivatives of $g_{\mu\nu}$ in the curvature scalar $R$.
To do that, one makes use of the following well-known formula which holds in general space-time dimensions: 
When one writes the scalar curvature
\begin{eqnarray}
R = R_1 +  R_2,
\label{R=R1+R2}
\end{eqnarray}
the formula takes the form \cite{Fujii}
\begin{eqnarray}
R_1 = -2 R_2 + \frac{1}{\sqrt{-g}} \partial_\mu (\sqrt{-g} A^\mu),
\label{Ident1}
\end{eqnarray}
where one has defined the following quantities
\begin{eqnarray}
R_1 &=& g^{\mu\nu} \left( \partial_\rho \Gamma^\rho_{\mu\nu} 
- \partial_\nu \Gamma^\rho_{\mu\rho} \right), \nonumber\\
R_2 &=& g^{\mu\nu} \left( \Gamma^\sigma_{\rho\sigma} \Gamma^\rho_{\mu\nu} 
- \Gamma^\sigma_{\rho\nu} \Gamma^\rho_{\mu\sigma} \right) \nonumber\\
&=& g^{\mu\nu} \Gamma^\sigma_{\rho\sigma} \Gamma^\rho_{\mu\nu} 
+ \frac{1}{2} \Gamma^\rho_{\mu\nu} \partial_\rho g^{\mu\nu}, \nonumber\\
A^\mu &=& g^{\nu\rho} \Gamma^\mu_{\nu\rho} - g^{\mu\nu} \Gamma^\rho_{\nu\rho}. 
\label{R&A}
\end{eqnarray}
Here let us note that $R_2$ is free of second-order derivatives of $g_{\mu\nu}$,
which are now involved in the term including $A^\mu$.
Using this formula, we can rewrite the Lagrangian density (\ref{WTDiff Lagr}) to the form
\begin{eqnarray}
{\cal L} = {\cal L}_0 + \frac{1}{2} \partial_\mu \left( |g|^{\frac{1}{n}} A^\mu \right),
\label{L0-1}
\end{eqnarray}
where ${\cal L}_0$ is defined as
\begin{eqnarray}
{\cal L}_0 = \frac{1}{2} |g|^{\frac{1}{n}} \left[ - R_2 + \frac{n-2}{2n} \frac{1}{|g|} A^\mu \partial_\mu |g|
+ \frac{(n-1)(n-2)}{4n^2} \frac{1}{|g|^2} g^{\mu\nu} \partial_\mu |g| \partial_\nu |g| \right].
\label{L0-2}
\end{eqnarray}

We are now ready to show that the Noether current for the Weyl symmetry is also zero by the more
conventional method. First of all, let us observe that the variation of ${\cal L}$ under the 
Weyl transformation (\ref{Weyl transf}) comes from only the total derivative term
\begin{eqnarray}
\delta {\cal L} = \partial_\mu \left[ (n-1) |g|^{\frac{1}{n}} g^{\mu\nu} \partial_\nu \Lambda \right]
= \frac{1}{2} \partial_\mu \left[ \delta( |g|^{\frac{1}{n}} A^\mu ) \right].
\label{delta L3}
\end{eqnarray}
The total derivative terms are irrelevant to dynamics so in what follows let us focus our attention only on
the Lagrangian ${\cal L}_0$, which is free of second-order derivatives of $g_{\mu\nu}$.

Second, by an explicit calculation we find that the Lagrangian ${\cal L}_0$ is invariant under the Weyl 
transformation without any surface term
\begin{eqnarray}
X_2^\mu = 0.
\label{X2}
\end{eqnarray}
Finally, applying the Noether theorem \cite{Noether} for ${\cal L}_0$, we can derive the following result
\begin{eqnarray}
K_2^\mu = \frac{\partial {\cal L}_0}{\partial ( \partial_\mu g_{\nu\rho} )} ( -2 g_{\nu\rho} ) = 0.
\label{K2}
\end{eqnarray}
Hence, the Noether current for the Weyl symmetry identically vanishes as before
\begin{eqnarray}
J_2^\mu = K_2^\mu - X_2^\mu = 0.
\label{J2}
\end{eqnarray}

At this stage, we should refer to an ambiguity associated with the Noether currents 
for local Weyl symmetry. Our calculation in this section is based on the Noether's first theorem, 
which is applicable for global symmetries, and the second theorem, which can be applied to local (gauge) 
symmetries. Of course, the latter case includes the former one as a special case, and 
the both Noether's theorems give the same result such that the Noether currents are identically vanishing. 
However, we should recall the well-known fact that the Noether currents for local (gauge) symmetries always 
reduce to superpotentials, which give us some ambiguity. Thus, the more reliable statement, 
which is obtained from our calculation at hand, is that the global Weyl symmetry has a vanishing Noether current, 
and hence neither charge nor symmetry generator.

Next, we shall provide a simpler proof that the Noether current for the Weyl symmetry in both the conformally invariant 
scalar-tensor gravity and WTDiff gravity vanishes. This proof is based on the observation that 
via the metric (\ref{Weyl invariant metric}) and the gauge condition (\ref{WTDiff gauge}),
the two theories become equivalent, and the Noether currents are gauge invariant quantities.
For simplicity, we will consider the action which includes only first-order derivatives of
the metric tensor $g_{\mu\nu}$.

As the starting action, we will take the action (\ref{Cof-inv S-T Action}) of the conformally invariant scalar-tensor 
gravity. As in the case of the WTDiff gravity, this action can be rewritten in the first-order derivative form
\begin{eqnarray}
S = \int d^n x \left[ {\cal L}_3 + \frac{n-2}{8(n-1)} \partial_\mu ( \sqrt{-g} \varphi^2 A^\mu ) \right],
\label{1st-Cof-inv S-T Action}
\end{eqnarray}
where ${\cal L}_3$ is defined by 
\begin{eqnarray}
{\cal L}_3 = \sqrt{-g} \left[ - \frac{n-2}{8(n-1)} \varphi^2 R_2 
- \frac{n-2}{8(n-1)} A^\mu \partial_\mu (\varphi^2)
+ \frac{1}{2} g^{\mu\nu} \partial_\mu \varphi \partial_\nu \varphi \right].
\label{L3}
\end{eqnarray}
The total derivative term in $S$ plays no role in bulk dynamics, so we will henceforth pay our attention
to ${\cal L}_3$. 
It is easy to show that ${\cal L}_3$ is invariant under the Weyl transformation without a surface term, which 
gives us
\begin{eqnarray}
X_3^\mu = 0.
\label{X_3}
\end{eqnarray}
Then, the Noether theorem \cite{Noether} provides us with
\begin{eqnarray}
K_3^\mu = \frac{\partial {\cal L}_3}{\partial ( \partial_\mu \varphi )} \frac{n-2}{2} \varphi
+ \frac{\partial {\cal L}_3}{\partial ( \partial_\mu g_{\nu\rho} )} ( -2 g_{\nu\rho} ).
\label{K_3}
\end{eqnarray}

Here we would like to give a simpler proof of $K_3^\mu = 0$ without much calculations.
The key observation for this proof is to recall that three kinds of gravitational theories are related to each other 
by a Weyl-invariant metric (\ref{Weyl invariant metric}), from which taking the differentiation, we can 
derive an equation
\begin{eqnarray}
\partial_\mu \hat g_{\nu\rho} = \left( \frac{1}{2} \sqrt{\frac{n-2}{n-1}} \varphi \right)^{\frac{4}{n-2}} 
\left( \frac{4}{n-2} \frac{1}{\varphi} \partial_\mu \varphi g_{\nu\rho} + \partial_\mu g_{\nu\rho} \right).
\label{Key}
\end{eqnarray}
Using this equation, one finds that
\begin{eqnarray}
\frac{\partial {\cal L}_3}{\partial ( \partial_\mu \varphi )}
&=& \frac{\partial {\cal L}_3}{\partial ( \partial_\mu \hat g_{\nu\rho} )} \left( \frac{1}{2} 
\sqrt{\frac{n-2}{n-1}} \varphi \right)^{\frac{4}{n-2}} \frac{4}{n-2} \frac{1}{\varphi} g_{\nu\rho},
\nonumber\\
\frac{\partial {\cal L}_3}{\partial ( \partial_\mu g_{\nu\rho} )}
&=& \frac{\partial {\cal L}_3}{\partial ( \partial_\mu \hat g_{\nu\rho} )} \left( \frac{1}{2} 
\sqrt{\frac{n-2}{n-1}} \varphi \right)^{\frac{4}{n-2}}.
\label{Key 2}
\end{eqnarray}
From Eq. (\ref{Key 2}), the equation (\ref{K_3}) produces the expected result
\begin{eqnarray}
K_3^\mu = 0.
\label{K_3-2}
\end{eqnarray}
As a result, the Noether current for the Weyl symmetry is vanishing
\begin{eqnarray}
J_3^\mu = K_3^\mu - X_3^\mu = 0.
\label{J_3}
\end{eqnarray}

This is our simpler proof of the vanishing Noether current for the Weyl symmetry in the conformally 
invariant scalar-tensor gravity. Since the current is gauge invariant, our proof can be directly 
applied to any conformally invariant gravitational theories such as the WTDiff gravity 
obtained via the trick (\ref{Weyl invariant metric}) and the gauge condition (\ref{WTDiff gauge}).
From our simple proof, we can also explain why the Weyl symmetry existing in both the conformally
invariant scalar-tensor gravity and the WTDiff gravity is identically vanishing. 
It has been already shown in the previous section 
that these two gravitational theories are equivalent to general relativity, and the Noether currents for
the Weyl symmetry are Weyl-invariant quantities. Since there is no Weyl symmetry in general relativity,
the Noether current for the Weyl symmetry should be trivially zero in general relativity. The equivalence 
among the three theories and the gauge invariance of the Noether currents naturally lead to a conclusion 
that the Weyl currents in the conformally invariant scalar-tensor gravity and the WTDiff gravity 
should be vanishing as well.     

So far, we have confined our attention to only the gravitational sector. Since there are plenty of matters 
around us, it is natural to ask if effects of matter fields could change our conclusion or not.  In the
following subsections, we will show that the introduction of conformal matters does not modify the fact 
that the Weyl current vanishes.

\subsection{Scalar field}

First, let us turn our attention to a real scalar field $\phi$ in an $n$-dimensional curved space-time. 
The action is consisted of a kinetic term and a potential $V(\phi)$
\begin{eqnarray}
S_\phi = \int d^n x |g|^{\frac{1}{2}} \left[ - \frac{1}{2} g^{\mu\nu} \partial_\mu \phi \partial_\nu \phi
- V(\phi)  \right].
\label{S-Action}
\end{eqnarray}
Note that this action is manifestly invariant under the full group of diffeomorphisms (Diff). 
Under the Weyl transformation, the scalar field $\phi$ has the same transformation law as the spurion 
field $\varphi$
\begin{eqnarray}
\phi \rightarrow \phi^\prime = \Omega^{-\frac{n-2}{2}}(x) \phi.
\label{S-Weyl transf}
\end{eqnarray}

The trick to enlarge gauge symmetries from Diff to WDiff is now to make a Weyl-invariant scalar field 
$\hat \phi  = \varphi^{-1} \phi$ in addition to the Weyl-invariant metric (\ref{Weyl invariant metric}), 
and then replace the metric and the scalar field in the action (\ref{S-Action}) by the corresponding 
Weyl-invariant objects. As a result, a WDiff-invariant scalar action takes the form
\begin{eqnarray}
\hat S_\phi &=& \int d^n x \ \hat {\cal L}_\phi                           \nonumber\\
&=& \int d^n x |\hat g|^{\frac{1}{2}} \left[ - \frac{1}{2} \hat g^{\mu\nu} \partial_\mu \hat \phi \partial_\nu \hat \phi
- V(\hat \phi)  \right]                                             \nonumber\\
&=& \int d^n x |g|^{\frac{1}{2}} \left[ - \frac{n-2}{8(n-1)} \varphi^2 g^{\mu\nu} 
\partial_\mu \left(\frac{\phi}{\varphi}\right) \partial_\nu \left(\frac{\phi}{\varphi}\right) 
- \left( \frac{1}{2} \sqrt{\frac{n-2}{n-1}} \varphi \right)^{\frac{2n}{n-2}} V\left(\frac{\phi}{\varphi}\right)  
\right].
\label{S-WDiff-Action}
\end{eqnarray}

We shall calculate the Noether current for Weyl symmetry by the two different methods. One method, which is called the WDiff method, 
is to calculate the current in the WDiff-invariant action without gauge-fixing the Weyl symmetry like the conformally invariant 
scalar-tensor gravity. The other method, which is called the WTDiff method, is to gauge-fix the longitudinal diffeomorphism 
by the gauge condition (\ref{WTDiff gauge}), by which the WDiff-invariant action is reduced to the WTDiff-invariant one, 
and then calculate the Noether current for the Weyl symmetry like the WTDiff gravity. The Weyl current is a gauge-invariant quantity, 
so both the methods should provide the same result.

First, let us calculate the Noether current for the Weyl symmetry on the basis of the WDiff matter action 
(\ref{S-WDiff-Action}). It is easy to see that the action (\ref{S-WDiff-Action}) is invariant under 
the Weyl transformation without a surface term, so we have
\begin{eqnarray}
X_\phi^\mu = 0.
\label{X_phi}
\end{eqnarray}
Again, the Noether theorem \cite{Noether} yields
\begin{eqnarray}
K_\phi^\mu = \frac{\partial \hat {\cal L}_\phi}{\partial ( \partial_\mu \phi )} \frac{n-2}{2} \phi
+ \frac{\partial \hat {\cal L}_\phi}{\partial ( \partial_\mu \varphi )} \frac{n-2}{2} \varphi
+ \frac{\partial \hat {\cal L}_\phi}{\partial ( \partial_\mu g_{\nu\rho} )} ( -2 g_{\nu\rho} ).
\label{K_phi}
\end{eqnarray}
Next, the Weyl-invariant combinations (\ref{Weyl invariant metric}) and
$\hat \phi  = \varphi^{-1} \phi$ give us the relations 
\begin{eqnarray}
\frac{\partial \hat {\cal L}_\phi}{\partial ( \partial_\mu \phi )}
&=& \frac{\partial \hat {\cal L}_\phi}{\partial ( \partial_\mu \hat \phi )} \frac{1}{\varphi},
\nonumber\\
\frac{\partial \hat {\cal L}_\phi}{\partial ( \partial_\mu \varphi )}
&=& \frac{\partial \hat {\cal L}_\phi}{\partial ( \partial_\mu \hat g_{\nu\rho} )} 
\left( \frac{1}{2} \sqrt{\frac{n-2}{n-1}} \varphi \right)^{\frac{4}{n-2}} \frac{4}{n-2} 
\frac{1}{\varphi} g_{\nu\rho}
- \frac{\partial \hat {\cal L}_\phi}{\partial ( \partial_\mu \hat \phi )} \frac{\phi}{\varphi^2},
\nonumber\\
\frac{\partial \hat {\cal L}_\phi}{\partial ( \partial_\mu g_{\nu\rho} )}
&=& \frac{\partial \hat {\cal L}_\phi}{\partial ( \partial_\mu \hat g_{\nu\rho} )} 
\left( \frac{1}{2} \sqrt{\frac{n-2}{n-1}} \varphi \right)^{\frac{4}{n-2}}.
\label{Key 3}
\end{eqnarray}
Using these relations (\ref{Key 3}), $K_\phi^\mu$ in (\ref{K_phi}) becomes zero 
\begin{eqnarray}
K_\phi^\mu = 0.
\label{K_phi-2}
\end{eqnarray}
The Noether current for the Weyl symmetry is therefore vanishing
\begin{eqnarray}
J_\phi^\mu = K_\phi^\mu - X_\phi^\mu = 0.
\label{J_phi}
\end{eqnarray}
This is a general result and even after gauge-fixing the longitudinal diffeomorphism 
this result should be valid since the Weyl current is gauge invariant under the
Weyl transformation. Indeed, this is so by calculating the Weyl current in WTDiff
scalar action below.

Now let us take the gauge condition (\ref{WTDiff gauge}) for the longitudinal diffeomorphism,
which does not break the local Weyl symmetry. Inserting the gauge condition (\ref{WTDiff gauge})
to the WDiff-invariant scalar action (\ref{S-WDiff-Action}) leads to the WTDiff-invariant scalar action
\begin{eqnarray}
\hat S_\phi &=& \int d^n x \ \hat {\cal L}_\phi                           \nonumber\\
&=& \int d^n x \Biggl\{ - \frac{n-2}{8(n-1)} |g|^{\frac{1}{2}} g^{\mu\nu} \left[ \partial_\mu \phi \partial_\nu \phi 
+ \frac{n-2}{2n} \frac{\phi}{|g|} \partial_\mu |g| \partial_\nu \phi   
+ \frac{(n-2)^2}{16 n^2} \frac{\phi^2}{|g|^2} \partial_\mu |g| \partial_\nu |g| \right]  \nonumber\\
&-& V\left(\frac{1}{2} \sqrt{\frac{n-2}{n-1}} |g|^{\frac{n-2}{4n}} \phi \right)  \Biggr\}.
\label{S-WTDiff-Action}
\end{eqnarray}
  
Since the action (\ref{S-WTDiff-Action}) is invariant under the Weyl transformation without a surface term, 
we have 
\begin{eqnarray}
X_\phi^\mu = 0.
\label{X_phi 2}
\end{eqnarray}
The Noether theorem \cite{Noether} gives us the formula
\begin{eqnarray}
K_\phi^\mu = \frac{\partial \hat {\cal L}_\phi}{\partial ( \partial_\mu \phi )} \frac{n-2}{2} \phi
+ \frac{\partial \hat {\cal L}_\phi}{\partial ( \partial_\mu g_{\nu\rho} )} ( -2 g_{\nu\rho} ).
\label{K_phi 2}
\end{eqnarray}
It is useful to evaluate each term in (\ref{K_phi 2}) separately to see its gauge invariance.
In fact, the result is given by 
\begin{eqnarray}
\frac{\partial \hat {\cal L}_\phi}{\partial ( \partial_\mu \phi )} \frac{n-2}{2} \phi
&=& - \frac{n-2}{4} \hat \phi^2 \hat g^{\mu\nu} \partial_\nu \log \left( \hat \phi^2 \right),   \nonumber\\
\frac{\partial \hat {\cal L}_\phi}{\partial ( \partial_\mu g_{\nu\rho} )} ( -2 g_{\nu\rho} )
&=& \frac{n-2}{4} \hat \phi^2 \hat g^{\mu\nu} \partial_\nu \log \left( \hat \phi^2 \right).
\label{K_phi 3}
\end{eqnarray}
As promised, each term is manifestly gauge invariant under the Weyl transformation since
it is expressed in terms of only gauge-invariant quantities. 
Adding the two terms in (\ref{K_phi 3}), we have
\begin{eqnarray}
K_\phi^\mu = 0.
\label{K_phi 4}
\end{eqnarray}
Thus, the Noether current for the Weyl symmetry in the WTDiff method is certainly vanishing
\begin{eqnarray}
J_\phi^\mu = K_\phi^\mu - X_\phi^\mu = 0.
\label{J_phi 2}
\end{eqnarray}
The both results in (\ref{J_phi}) and (\ref{J_phi 2}) clearly account for that the Noether current
for the Weyl symmetry is vanishing in the both WDiff-invariant scalar action and WTDiff-invariant one.

\subsection{Vector field}

Next, we will move on to spin 1 abelian gauge field, that is, the electro-magnetic field. It is well-known
that the Maxwell action for the electro-magnetic field is invariant in only four space-time dimensions,
but not so in an arbitrary space-time dimension. It is therefore necessary to extend the Maxwell action
in four dimensions in such a way it is also invariant under the Weyl transformation in general dimensions.
We are now accustomed to the recipe for accomplishing this work: Start with a Diff-invariant action and
then replace all fields with the corresponding Weyl-invariant fields, by which we have the WDiff-invariant action. 
Furthermore, the WTDiff-invariant action is obtained by selecting the gauge condition (\ref{WTDiff gauge}) 
for the longitudinal diffeomorphism. According to this recipe, let us start with the conventional Maxwell action
which is invariant under Diff in $n$ space-time dimensions:
\begin{eqnarray}
S_A = - \frac{1}{4} \int d^n x |g|^{\frac{1}{2}} g^{\mu\nu} g^{\rho\sigma} F_{\mu\rho} F_{\nu\sigma},
\label{A-Action}
\end{eqnarray}
where $F_{\mu\nu} = \partial_\mu A_\nu - \partial_\nu A_\mu$. The Weyl transformation for the vector field
is defined as usual
\begin{eqnarray}
A_\mu \rightarrow A^\prime_\mu = A_\mu.
\label{Weyl transf for A}
\end{eqnarray}
Then, the WDiff-invariant action reads
\begin{eqnarray}
\hat S_A &=& \int d^n x \hat {\cal L}_A     \nonumber\\
&=& - \frac{1}{4} \int d^n x |\hat g|^{\frac{1}{2}} \hat g^{\mu\nu} \hat g^{\rho\sigma} F_{\mu\rho} F_{\nu\sigma},
\nonumber\\
&=& - \frac{1}{4} \int d^n x |g|^{\frac{1}{2}} \left( \frac{1}{2} \sqrt{\frac{n-2}{n-1}} 
\varphi \right)^{\frac{2(n-4)}{n-2}} g^{\mu\nu} g^{\rho\sigma} F_{\mu\rho} F_{\nu\sigma},
\label{WDiff-A-Action}
\end{eqnarray}
and the WTDiff-invariant action takes the form
\begin{eqnarray}
\hat S_A &=& \int d^n x \hat {\cal L}_A     \nonumber\\
&=& - \frac{1}{4} \int d^n x |g|^{\frac{2}{n}} g^{\mu\nu} g^{\rho\sigma} F_{\mu\rho} F_{\nu\sigma}.
\label{WTDiff-A-Action}
\end{eqnarray}

Based on these actions, it is again easy to evaluate the Noether current associated with the Weyl symmetry
by the two methods. For instance, in the WTDiff method, since the WTDiff-invariant action is invariant 
without a surface term, we have 
\begin{eqnarray}
X_A^\mu = 0.
\label{X_A}
\end{eqnarray}
The Noether theorem \cite{Noether} again produces the formula
\begin{eqnarray}
\Lambda K_A^\mu = \frac{\partial \hat {\cal L}_A}{\partial ( \partial_\mu A_\nu )} \delta A_\nu
+ \frac{\partial \hat {\cal L}_A}{\partial ( \partial_\mu g_{\nu\rho} )} \delta g_{\nu\rho},
\label{K_A}
\end{eqnarray}
where $\Lambda$ is the infinitesimal parameter for the Weyl transformation. 
Since $\delta A_\nu = \frac{\partial \hat {\cal L}_A}{\partial ( \partial_\mu g_{\nu\rho} )} = 0$,
we soon reach the result
\begin{eqnarray}
K_A^\mu = 0.
\label{K_A2}
\end{eqnarray}
Hence, we have the vanishing Noether current
\begin{eqnarray}
J_A^\mu = K_A^\mu - X_A^\mu = 0.
\label{J_A}
\end{eqnarray}
It is straightforward to derive the same result on the basis of the WDiff-invariant action 
(\ref{WDiff-A-Action}).

\subsection{Spinor field}

Finally, as one of matter fields, let us consider the Dirac spinor field. It is known that in general $n$
space-time dimensions, the action for massless Dirac spinor fields is invariant under the Weyl transformation
\cite{Englert}.  We find it useful to recall the symmetry properties of the Dirac action whose
Lagrangian density is \footnote{See Appendix A.2 for notation and some definitions related to spinors.}
\begin{eqnarray}
{\cal {L}}_\psi  &=& - \frac{1}{2} e \bar \psi \left( \slashb D - \overleftarrow {\slashb D} \right) \psi
- e m \bar \psi \psi      \nonumber\\
&=& e \left( - \bar \psi e_a ^\mu \gamma^a D_\mu \psi - m \bar \psi \psi \right),
\label{Dirac action}
\end{eqnarray}
where $e = \det e_{a \mu}$, $\slashb D = \gamma^\mu D_\mu$, and in the last equality we have used the
integration by parts. In case of the massless Dirac field ($m = 0$), in general $n$ space-time dimensions,
the action $\int d^n x {\cal {L}}_\psi$ is invariant under the following Weyl transformation
\begin{eqnarray}
e_\mu ^a \rightarrow e_\mu ^{\prime \, a} = \Omega(x) e_\mu ^a, \quad
e^{a \mu} \rightarrow e^{\prime \, a \mu} = \Omega^{-1}(x) e^{a \mu}, \quad
\psi \rightarrow \psi^\prime = \Omega^{-\frac{n-1}{2}}(x) \psi.
\label{Spinor Weyl transf}
\end{eqnarray}
 
With the presence of the scalar field $\varphi$, we can make even the mass term be invariant under the Weyl transformation.
To do that, as before, it is sufficient to introduce the Weyl invariant fields and then replace each field in the
Lagrangian (\ref{Dirac action}) by the corresponding Weyl invariant field.  The Weyl invariant fields are given by
\begin{eqnarray}
\hat e_a ^\mu = \left( \frac{1}{2} \sqrt{\frac{n-2}{n-1}} \varphi \right)^{- \frac{2}{n-2}} e_a ^\mu, \quad
\hat \psi =  \left( \frac{1}{2} \sqrt{\frac{n-2}{n-1}} \varphi \right)^{- \frac{n-1}{n-2}} \psi.
\label{Spinor Weyl-inv fields}
\end{eqnarray}
By replacing each field with the corresponding Weyl invariant one in Eq. (\ref{Dirac action}), we have
the Weyl invariant massive Dirac Lagrangian density $\hat {\cal {L}}_\psi$
\begin{eqnarray}
\hat {\cal {L}}_\psi  
&=& \hat e \left( - \hat{\bar \psi} \hat e_a ^\mu \gamma^a \hat D_\mu \hat \psi - m \hat{\bar \psi} \hat \psi \right),
\nonumber\\
&=& e \left[ - \bar \psi e_a ^\mu \gamma^a D_\mu \psi 
- \left( \frac{1}{2} \sqrt{\frac{n-2}{n-1}} \varphi \right)^{\frac{2}{n-2}} m \bar \psi \psi \right].
\label{Weyl-inv Dirac action}
\end{eqnarray}
Let us note that the first term has the same expression as before since the massless Dirac Lagrangian density is 
invariant under the Weyl transformation. The action $\int d^n x \hat {\cal {L}}_\psi$ is invariant under both the Weyl 
transformation and the full group of Diff. To reduce the symmetries from WDiff to WTDiff, we will take the
gauge condition (\ref{WTDiff gauge}) for the longitudinal diffeomorphism. The resulting Lagrangian density reads
\begin{eqnarray}
\hat {\cal {L}}_\psi  
= e \left( - \bar \psi e_a ^\mu \gamma^a D_\mu \psi - e^{-\frac{1}{n}} m \bar \psi \psi \right).
\label{WTDiff Dirac action}
\end{eqnarray}

The Noether current for the Weyl symmetry should be calculated by using either action since the current is a gauge
invariant quantity. We will use the Lagranagian density (\ref{WTDiff Dirac action}), which is invariant
under the Weyl transformation without surface terms, i.e. 
\begin{eqnarray}
X_\psi ^\mu = 0.
\label{X_psi}
\end{eqnarray}
The Noether theorem gives us the expression 
\begin{eqnarray}
\Lambda K_\psi ^\mu = \frac{\partial^R \hat {\cal {L}}_\psi}{\partial (\partial_\mu \psi)} \delta \psi
+ \frac{\partial \hat {\cal {L}}_\psi}{\partial (\partial_\mu e^a _\nu)} \delta e^a _\nu,
\label{K_psi}
\end{eqnarray}
where we have used the right-derivative notation with respect to the spinor field and the second-order
formalism of gravity, that is, the spin connection has been regarded as a function of the vielbein.
A straightforward calculation of each term in (\ref{K_psi}) yields
\begin{eqnarray}
\frac{\partial^R \hat {\cal {L}}_\psi}{\partial (\partial_\mu \psi)} \delta \psi
&=& - \Lambda \frac{n-1}{2} e \bar \psi e_a ^\mu \gamma^a \psi 
\nonumber\\
\frac{\partial \hat {\cal {L}}_\psi}{\partial (\partial_\mu e^a _\nu)} \delta e^a _\nu
&=& \Lambda \frac{n-1}{2} e \bar \psi e_a ^\mu \gamma^a \psi,
\label{Each K_psi}
\end{eqnarray}
both of which are gauge invariant as expected. Therefore, we have
\begin{eqnarray}
K_\psi ^\mu = 0.
\label{Zero K_psi}
\end{eqnarray}
Accordingly, even in this case, we have the identically vanishing Noether current
\begin{eqnarray}
J_\psi ^\mu = K_\psi ^\mu - X_\psi ^\mu = 0.
\label{J_psi}
\end{eqnarray}

To close this section, we should comment on the trace or Weyl (conformal) anomaly. 
It is well-known that in a curved space-time, certain matter fields, such as the electro-magnetic 
field in four dimensions and massless Dirac fields in any dimensions, exhibits Weyl (local conformal) 
invariance at the classical level as mentioned above. The Weyl invariance of the action implies that
the trace of the energy-momentum tensor is zero. We are also familiar with the fact that a theory 
based on a classical action which is Weyl invariant in general loses its Weyl invariance in the quantum
theory as a result of renormalization, i.e. owing to the existence of the renormalization scale. 
The energy-momentum tensor therefore acquires a non-zero trace, known as the trace or Weyl (conformal) 
anomaly \cite{Capper1, Capper2}. 

However, this well-known result does not generally hold in the present formalism where there is 
the spurion field $\varphi$. In our formalism, we keep the situation in mind such that the conformally 
invariant scalar-tensor gravity coexists with the various conformally invariant matter fields. In this
situation, the spurion field $\varphi$ is assumed to be broken spontaneouely $\varphi = \langle \varphi \rangle
+ \sigma$ where the massless "meson" $\sigma$ is the Goldstone boson restoring conformal symmetry,
even if there is no potential for triggering the spontaneous symmetry breakdown. Note that $\sigma = 0$
corresponds to the "unitary gauge" leading to general relativity or the WTDiff gravity depending on the
choice of $\langle \varphi \rangle$. The key idea is that we can use the vacuum expectation value of the spurion field, 
$\langle \varphi \rangle$, as the renormalization scale instead of the conventional fixed renormalization 
scale $\mu$ \cite{Englert}-\cite{Ghilencea2}. With this idea, we have a conformally invariant effective potential
without trace anomaly and the coupling constants still run with the momentum scale \cite{Tamarit}.

\section{Schwarzschild solution}

In this section, we wish to show that the Schwarzschild metric is a classical solution to the equations
of motion of the WTDiff gravity, Eq. (\ref{Eq from WTDiff Action}), or equivalently
Eq. (\ref{Cov-Eq from WTDiff Action}). Before doing so, we soon realize that a notable feature of 
Eq. (\ref{Eq from WTDiff Action}) is that the traceless Einstein tensor defined as 
$G_{\mu\nu}^T = R_{\mu\nu} - \frac{1}{n} g_{\mu\nu} R$ in the LHS has a beautiful geometrical structure 
whereas the traceless energy-momentum tensor $T_{(g)\mu\nu}^T = T_{(g)\mu\nu} - \frac{1}{n} g_{\mu\nu} T_{(g)}$ 
in the RHS has a complicated expression, and the presence of the metric determinant $g$ and 
its derivative $\nabla_\mu \nabla_\nu |g|$ reflects the fact that the equations of motion are not invariant 
under Diff, but only TDiff. It is therefore natural to fix the Weyl symmetry first by the gauge condition 
\begin{eqnarray}
g = -1,
\label{g=-1}
\end{eqnarray}
which is nothing but the unimodular condition (\ref{Unimodularity}). 
Since the traceless energy-momentum tensor in the RHS of Eq. (\ref{Eq from WTDiff Action}) trivially vanishes,
the resultant equations of motion read
\begin{eqnarray}
G_{\mu\nu}^T \equiv R_{\mu\nu} - \frac{1}{n} g_{\mu\nu} R = 0.
\label{Einstein spaces}
\end{eqnarray}
The space-time defined by Eq. (\ref{Einstein spaces}) is called Einstein spaces in four dimensions
and the study of the Riemannian spaces which are conformally related to Einstein spaces, has been
addressed for a long time \cite{Kozameh}. 

Now we wish to show that the Schwarzschild metric in the Cartesian coordinate system is a classical solution 
to the equations of motion (\ref{Einstein spaces}).
For this purpose, we will look for a gravitational field outside an isolated, static, spherically symmetric object
with mass $M$. In the far region from the isolated object, we assume that the metric tensor is in an asymptotically
Lorentzian form
\begin{eqnarray}
g_{\mu\nu} \rightarrow \eta_{\mu\nu} + {\cal {O}}\left(\frac{1}{r^{n-3}}\right),
\label{BC}
\end{eqnarray}
where $\eta_{\mu\nu}$ is the Minkowski metric, and the radial coordinate $r$ is defined as 
\begin{eqnarray}
r = \sqrt{(x^1)^2 + (x^2)^2 + \cdots + (x^{n-1})^2} = \sqrt{(x^i)^2},
\label{radial}
\end{eqnarray}
with $i$ running over spatial coordinates ($i = 1, 2, \cdots, n-1$).

Let us recall that the most spherically symmetric line element in $n$ space-time dimensions 
reads
\begin{eqnarray}
d s^2 = - A(r) d t^2 + B(r) (x^i d x^i)^2  + C(r) (d x^i)^2 + D(r) d t \ x^i d x^i,
\label{Line element 1}
\end{eqnarray}
where $A(r)$ and $C(r)$ are positive functions depending on only $r$. Requiring
the invariance under the time reversal $t \rightarrow -t$ leads to $D = 0$. As is well-known,
we can set $C(r) = 1$ by redefining the radial coordinate $r$ \cite{Adler}. Thus, the line
element under consideration takes the form in the Cartesian coordinate system
\begin{eqnarray}
d s^2 = - A(r) d t^2 + (d x^i)^2 + B(r) (x^i d x^i)^2.
\label{Line element 2}
\end{eqnarray}

From this line element (\ref{Line element 2}), the non-vanishing components of the metric tensor 
are given by
\begin{eqnarray}
g_{tt} = - A, \quad  g_{ij} = \delta_{ij} + B x^i x^j,
\label{Metric}
\end{eqnarray}
and the components of its inverse matrix are 
\begin{eqnarray}
g^{tt} = - \frac{1}{A}, \quad  g^{ij} = \delta^{ij} - \frac{B}{1 + B r^2} x^i x^j.
\label{Inverse Metric}
\end{eqnarray}
Moreover, using these components of the metric tensor, the affine connection is calculated to be
\begin{eqnarray}
\Gamma^t_{ti} &=& \frac{A^\prime}{2A} \frac{x^i}{r}, \quad 
\Gamma^i_{tt} = \frac{A^\prime}{2(1 + B r^2)} \frac{x^i}{r}, \nonumber\\ 
\Gamma^i_{jk} &=& \frac{1}{2(1 + B r^2)} \frac{x^i}{r} ( 2B r \delta_{jk} + B^\prime x^j x^k ),
\label{Affine}
\end{eqnarray}
where the prime denotes the differentiation with respect to $r$, for instance, $A^\prime = \frac{dA}{dr}$. 

Here let us take the gauge condition (\ref{g=-1}) for the Weyl transformation into consideration. 
By means of the metric tensor (\ref{Metric}), the gauge condition (\ref{g=-1}) is cast to the form
\begin{eqnarray}
A (1 + B r^2) = 1.
\label{AB}
\end{eqnarray}
Using this gauge condition (\ref{AB}) and Eqs. (\ref{Metric})-(\ref{Affine}), the Ricci tensor
and the scalar curvature can be easily calculated to be  
\begin{eqnarray}
R_{tt} &=& \frac{1}{2} A ( A^{\prime\prime} + \frac{n-2}{r} A^\prime ), \nonumber\\ 
R_{ij} &=& \left[ \frac{n-3}{r^2} ( 1 - A ) - \frac{A^\prime}{r} \right] \delta_{ij}
+ \frac{1}{r^2} \Biggl[ \frac{n-3}{r^2} ( A - 1 ) 
+ \frac{1}{r} \frac{A^\prime}{A} ( 1 - \frac{n}{2} + A ) - \frac{1}{2} \frac{A^{\prime\prime}}{A} \Biggr] 
x^i x^j,
\nonumber\\
R &=& - A^{\prime\prime} - \frac{2(n-2)}{r} A^\prime - \frac{(n-2)(n-3)}{r^2} ( A - 1 ). 
\label{Curvature}
\end{eqnarray}
These results produce the concrete expressions for the non-vanishing components of the traceless Einstein 
tensor $G_{\mu\nu}^T \equiv R_{\mu\nu} - \frac{1}{n} g_{\mu\nu} R$
\begin{eqnarray}
G_{tt}^T &=& \left( \frac{1}{2} - \frac{1}{n} \right) A \left[ A^{\prime\prime} + (n-4) \frac{1}{r} A^\prime 
- 2 (n-3) \frac{1}{r^2} ( A - 1 ) \right], \nonumber\\ 
G_{ij}^T &=& \left\{ \frac{1}{n} \delta_{ij} + \frac{1}{r^2} \frac{1}{A} \left[ -\frac{1}{2} + \frac{1}{n}( 1 - A ) \right]
x^i x^j \right\} \Biggl[ A^{\prime\prime}  + (n-4) \frac{1}{r} A^\prime 
\nonumber\\
&-& 2 (n-3) \frac{1}{r^2} ( A - 1 ) \Biggr].
\label{G^T}
\end{eqnarray}

Consequently, Eq. (\ref{Einstein spaces}) reduces to an equation
\begin{eqnarray}
A^{\prime\prime} + (n-4) \frac{1}{r} A^\prime - 2 (n-3) \frac{1}{r^2} ( A - 1 ) = 0.
\label{A-eq}
\end{eqnarray}
Noticing that the LHS of Eq. (\ref{A-eq}) can be rewritten as 
\begin{eqnarray}
&{}& A^{\prime\prime} + (n-4) \frac{1}{r} A^\prime - 2 (n-3) \frac{1}{r^2} ( A - 1 ) 
\nonumber\\
&=& \frac{1}{r^{n-3}} \frac{d^2}{d r^2} \left[ r^{n-3} ( A - 1 ) \right]
- (n-2) \frac{1}{r^{n-2}} \frac{d}{d r} \left[ r^{n-3} ( A - 1 ) \right],
\label{A-eq 2}
\end{eqnarray}
Eq. (\ref{A-eq}) is easily solved to be
\begin{eqnarray}
A(r) = 1 - \frac{2M}{r^{n-3}} + a r^2,
\label{A-value}
\end{eqnarray}
where $M$ and $a$ are integration constants. From the boundary condition (\ref{BC}),
we have to choose $a = 0$, and we can obtain the expression for $B(r)$ in terms of the gauge
condition (\ref{AB}). Accordingly, we arrive at the expressions for $A(r)$ and $B(r)$ 
\begin{eqnarray}
A(r) = 1 - \frac{2M}{r^{n-3}}, \quad B(r) = \frac{2M}{r^2 (r^{n-3} - 2M)}.
\label{AB-value}
\end{eqnarray}
Then, the line element is of form
\begin{eqnarray}
d s^2 = - \left( 1 - \frac{2M}{r^{n-3}} \right) d t^2 + (d x^i)^2 
+ \frac{2M}{r^2 (r^{n-3} - 2M)} (x^i d x^i)^2.
\label{Schwarzschild 1}
\end{eqnarray}
In this way, we have succeeded in showing that the Schwarzschild metric in the Cartesian coordinate system
is a classical solution in the WTDiff gravity as in general relativity.

However, there is a caveat. The Schwarzschild metric in the Cartesian coordinate system, 
(\ref{Schwarzschild 1}) can be rewritten in the spherical coordinate system as
\begin{eqnarray}
d s^2 = - \left( 1 - \frac{2M}{r^{n-3}} \right) d t^2 
+ \frac{1}{1 - \frac{2M}{r^{n-3}}} d r^2 + r^2 d \Omega_{n-2}^2,
\label{Schwarzschild 2}
\end{eqnarray}
where 
\begin{eqnarray}
d \Omega_{n-2}^2 = d \theta_2^2 + \sin^2 \theta_2 d \theta_3^2 + \cdots 
+ \prod_{i=2}^{n-2} \sin^2 \theta_i d \theta_{n-1}^2.
\label{Omega}
\end{eqnarray}
This form of the Schwarzschild metric is very familiar with physicists, but this is not a classical 
solution to the equations of motion of the WTDiff gravity, (\ref{Einstein spaces}). The reason 
is that when transforming from the Cartesian coordinates to the spherical coordinates, we have
a non-vanishing Jacobian factor which is against TDiff. To put differently, while the determinant of the
metric tensor in Eq. (\ref{Schwarzschild 1}) is $-1$, the one in Eq. (\ref{Schwarzschild 2}) is not so,
which is against the gauge condition (\ref{g=-1}). In order to show that Eq. (\ref{Schwarzschild 2}) is
also a classical solution, one has to solve the equations of motion under the condition $g \not= -1$,
which is at present a difficult task due to the complicated structure of the energy-momentum tensor.

\section{Charged black hole solution}

In the previous section, we have investigated classical solutions in the WTDiff gravity and found that
the Schwarzschild metric is indeed a classical solution to the equations of motion of the WTDiff gravity. 
A study of the Schwarzschild solution is of physical importance since the Schwarzschild solution 
corresponds to the basic one-body problem of classical astronomy, and the reliable experimental verifications 
of the Einstein equations are almost based on the Schwarzschild line element.
Then, it is natural to ask ourselves whether a charged black hole metric is also a classical solution 
to the equations of motion of the WTDiff gravity coupled to an electro-magnetic field or not.
In this section, we will prove that it is indeed the case in general $n$ space-time dimensions.

Our starting action is the sum of the WTDiff gravity action (\ref{WTDiff Action}) and the WTDiff-invariant Maxwell 
action (\ref{WTDiff-A-Action}) 
\begin{eqnarray}
S = \int d^n x \left\{ \frac{1}{2} |g|^{\frac{1}{n}} \left[ R + \frac{(n-1)(n-2)}{4n^2} \frac{1}{|g|^2}
g^{\mu\nu} \partial_\mu |g| \partial_\nu |g|  \right] 
- \frac{1}{4} |g|^{\frac{2}{n}} g^{\mu\nu} g^{\rho\sigma} F_{\mu\rho} F_{\nu\sigma} \right\}.
\label{WTDiff+EM Action}
\end{eqnarray}
It is worthwhile to point out that although the WTDiff gravity has been already shown to be equivalent to
general relativity, the WTDiff-invariant Maxwell action for the vector field $A_\mu$ is not equivalent to the
conventional Maxwell action except in four dimensions and is its Weyl-invariant generalization. Thus, it is
a nontrivial task to examine whether the action (\ref{WTDiff+EM Action}) possesses a charged black hole solution 
as a classical solution.
   
It is straightforward to derive the equations of motion for the gauge field $A_\mu$ and the metric tensor $g_{\mu\nu}$. 
The result is given by
\begin{eqnarray}
\partial_\mu ( |g|^{\frac{2}{n}} F^{\mu\nu} ) = 0,
\label{Maxwell eq of WTDiff+EM}
\end{eqnarray}
and
\begin{eqnarray}
R_{\mu\nu} - \frac{1}{n} g_{\mu\nu} R = T_{(g,A)\mu\nu} - \frac{1}{n} g_{\mu\nu} T_{(g,A)},
\label{Metric eq of WTDiff plus EM}
\end{eqnarray}
where the energy-momentum tensor $T_{(g,A)\mu\nu}$ is defined as
\begin{eqnarray}
T_{(g,A)\mu\nu} = \frac{(n-2)(2n-1)}{4n^2} \frac{1}{|g|^2} \partial_\mu |g| \partial_\nu |g|
- \frac{n-2}{2n} \frac{1}{|g|} \nabla_\mu \nabla_\nu |g| + |g|^{\frac{1}{n}} F_{\mu\alpha}
F_\nu \, ^\alpha.
\label{T_{(g,A)}}
\end{eqnarray}

This energy-momentum tensor $T_{(g,A)\mu\nu}$ is not covariantly conserved as $T_{(g)\mu\nu}$ in Eq. (\ref{T(g)}),
but it is possible to construct a covariantly conserved energy-momentum tensor as before. Along the same line of
arguments as in Section 2, the covariantly conserved energy-momentum tensor is found to be
\begin{eqnarray}
T_{\mu\nu} = T_{(g,A)\mu\nu} + \frac{n-2}{2n} g_{\mu\nu} \left[ - \frac{5n-3}{4n} \frac{1}{|g|^2} 
(\partial_\rho |g|)^2 + \frac{1}{|g|} \nabla_\rho \nabla^\rho |g| \right]
- \frac{1}{4} |g|^{\frac{1}{n}} g_{\mu\nu} F_{\rho\sigma} F^{\rho\sigma}.
\label{new T}
\end{eqnarray}
Moreover, as expected from the Weyl invariance of the WTDiff-invariant Maxwell action, this covariantly conserved
energy-momentum tensor $T_{\mu\nu}$ satisfies the relation
\begin{eqnarray}
T_{\mu\nu} - \frac{1}{n} g_{\mu\nu} T = T_{(g,A)\mu\nu} - \frac{1}{n} g_{\mu\nu} T_{(g,A)}.
\label{EM-Critical relation of T}
\end{eqnarray}
Hence, as in the case of the absence of the electro-magnetic field, even if we add the electro-magnetic field to
the WTDiff gravity, we can derive the standard Einstein equations (\ref{Einstein eq}) where the cosmological
constant appears as an integration constant. 

Since we want to find a charged black hole solution, we look for a gravitational field outside an isolated, static, 
spherically symmetric object with mass $M$ and electric charge $Q$. We again take the asymptotically Lorentzian
space-time, Eq. (\ref{BC}), as a boundary condition for the metric tensor. 
We also work with the line element (\ref{Line element 2}), and take the unimodular condition (\ref{g=-1}) 
as a gauge condition for the Weyl symmetry, so in this case we have perfectly the same equations as Eqs. 
(\ref{Line element 2})-(\ref{G^T}).
With the unimodular condition (\ref{g=-1}) for the Weyl symmetry, the Maxwell equation and the energy-momentum tensor
are respectively reduced to the form
\begin{eqnarray}
\partial_\mu F^{\mu\nu} &=& 0, 
\label{T & F 1} \\
T_{(g,A)\mu\nu} &=& F_{\mu\alpha} F_\nu \, ^\alpha.
\label{T & F 2}
\end{eqnarray}

As for the electro-magnetic field $A_\mu(x)$, we assume that it has a static, spherically symmetric
electric potential
\begin{eqnarray}
A_t = - \phi (r), \quad  A_i = 0,
\label{Gauge field}
\end{eqnarray}
where $\phi (r)$ is a function of $r$. 
First, let us solve the Maxwell equation (\ref{T & F 1}). With the ansatzes (\ref{Line element 2})
and (\ref{Gauge field}), the Maxwell equation (\ref{T & F 1}) is cast to a single equation
\begin{eqnarray}
\frac{d}{dr} ( r^{n-2} \phi^\prime ) = 0,
\label{Maxwell eq of WTDiff+EM 2}
\end{eqnarray}
which is easily integrated to be
\begin{eqnarray}
\phi(r) = \sqrt{\frac{n-2}{n-3}} \frac{Q}{r^{n-3}} + c,
\label{phi-sol}
\end{eqnarray}
where $Q$, which corresponds to an electric charge, and $c$ are integration constants. To fix the constant
$c$, we will impose a boundary condition
\begin{eqnarray}
\lim_{r \rightarrow \infty} \phi(r) = 0,
\label{BC for phi}
\end{eqnarray}
which uniquely determines $c = 0$. Thus, we obtain the final expression for $\phi(r)$
\begin{eqnarray}
\phi(r) = \sqrt{\frac{n-2}{n-3}} \frac{Q}{r^{n-3}}.
\label{phi-sol 2}
\end{eqnarray}

Next, let us try to solve the {\it{traceless}} Einstein equations (\ref{Metric eq of WTDiff plus EM})
with the unimodular gauge condition (\ref{g=-1}). 
For this purpose, we will calculate the {\it{traceless}} energy-momentum tensor defined as 
$T_{(g,A)\mu\nu}^T \equiv T_{(g,A)\mu\nu} - \frac{1}{n} g_{\mu\nu} T_{(g,A)}$ whose result is summarized as
\begin{eqnarray}
T_{(g,A)tt}^T &=& A \frac{(n-2)^2 (n-3)}{n} \frac{Q^2}{r^{2(n-2)}}, \nonumber\\ 
T_{(g,A)ij}^T &=& \frac{2(n-2)(n-3)}{n} \left( \delta_{ij} - \frac{A + \frac{n}{2} - 1}{A} \frac{x^i x^j}{r^2}  
\right) \frac{Q^2}{r^{2(n-2)}}. 
\label{T_{(g,A)}^T}
\end{eqnarray}
Consequently, the {\it{traceless}} Einstein equations (\ref{Metric eq of WTDiff plus EM}) reduce to 
an equation
\begin{eqnarray}
A^{\prime\prime} + (n-4) \frac{1}{r} A^\prime - 2 (n-3) \frac{1}{r^2} ( A - 1 ) 
- 2 (n-2)(n-3) \frac{Q^2}{r^{2(n-2)}} = 0.
\label{EM-A-eq}
\end{eqnarray}
This equation can be rewritten as 
\begin{eqnarray}
\frac{1}{r^{n-3}} \frac{d^2}{d r^2} \left[ r^{n-3} \left( A - 1 - \frac{Q^2}{r^{2(n-3)}} \right) \right]
- (n-2) \frac{1}{r^{n-2}} \frac{d}{d r} \left[ r^{n-3} \left( A - 1 - \frac{Q^2}{r^{2(n-3)}} \right) \right] = 0.
\label{EM-A-eq 2}
\end{eqnarray}
By performing an integration, $A(r)$ turns out to be
\begin{eqnarray}
A(r) = 1 - \frac{2M}{r^{n-3}} + \frac{Q^2}{r^{2(n-3)}} + a r^2,
\label{EM-A-value}
\end{eqnarray}
where $M$ and $a$ are integration constants. From the boundary condition (\ref{BC}),
we have to choose $a = 0$, and we can obtain the expression for $B(r)$ in terms of the gauge
condition (\ref{AB}). As a result, we reach the expressions for $A(r)$ and $B(r)$ 
\begin{eqnarray}
A(r) = 1 - \frac{2M}{r^{n-3}} + \frac{Q^2}{r^{2(n-3)}}, 
\quad B(r) = \frac{2Mr^{n-3} - Q^2}{r^2 \left[r^{2(n-3)} - 2Mr^{n-3} + Q^2 \right]}.
\label{EM-AB-value}
\end{eqnarray}
Then, the line element is of form
\begin{eqnarray}
d s^2 = - \left[ 1 - \frac{2M}{r^{n-3}} + \frac{Q^2}{r^{2(n-3)}} \right] d t^2 + (d x^i)^2 
+ \frac{2Mr^{n-3} - Q^2}{r^2 \left[r^{2(n-3)} - 2Mr^{n-3} + Q^2 \right]} (x^i d x^i)^2.
\label{CBH 1}
\end{eqnarray}
Hence, we have shown that the charged black hole metric in the Cartesian coordinate system
is indeed a classical solution in the WTDiff gravity coupled to the WTDiff-invariant Maxwell theory in an arbitrary
space-time dimension.

Again we should refer to an important remark. The charged black hole metric (\ref{CBH 1}) in the Cartesian coordinate 
system can be rewritten in a more familiar form in the spherical coordinate system
\begin{eqnarray}
d s^2 = - \left[ 1 - \frac{2M}{r^{n-3}} + \frac{Q^2}{r^{2(n-3)}} \right] d t^2 
+ \frac{1}{1 - \frac{2M}{r^{n-3}} + \frac{Q^2}{r^{2(n-3)}} } d r^2 + r^2 d \Omega_{n-2}^2.
\label{CBH 2}
\end{eqnarray}
However, this expression (\ref{CBH 2}) is {\it{not}} a classical solution in the WTDiff gravity plus 
the WTDiff-invariant Maxwell theory. This situation is very similar to the Schwarzschild black hole meric.
Namely, the dependence of classical solutions on the chosen coordinate system is a notable feature of the WTDiff gravity 
where there is no the full group of diffeomorphisms, but only TDiff.

\section{Cosmology}
 
As a final application of the classical WTDiff gravity, we would like to consider cosmology in the WTDiff gravity  
coupled with the WTDiff-invariant scalar matter. Before attempting to solve the {\it{traceless}} Einstein equations,
following the same method as before we can construct the energy-momentum tensor satisfying the covariant conservation 
law in this case as well. An interesting point here is that such a covariantly conserved energy-momentum tensor plays a
critical role in the construction of classical solutions, which should be contrasted to the cases treated thus far
where the covariantly conserved energy-momentum tensors are only needed to have a connection with the standard
Einstein equations.

The action with which we begin is the sum of the WTDiff gravity action (\ref{WTDiff Action}) and the WTDiff-invariant 
scalar action (\ref{S-WTDiff-Action}) 
\begin{eqnarray}
S &=& \int d^n x \Biggl\{ \frac{1}{2} |g|^{\frac{1}{n}} \left[ R + \frac{(n-1)(n-2)}{4n^2} \frac{1}{|g|^2}
g^{\mu\nu} \partial_\mu |g| \partial_\nu |g|  \right] 
- \frac{n-2}{8(n-1)} |g|^{\frac{1}{2}} g^{\mu\nu} \Biggl[ \partial_\mu \phi \partial_\nu \phi
\nonumber\\
&+& \frac{n-2}{2n} \frac{\phi}{|g|} \partial_\mu |g| \partial_\nu \phi
+ \frac{(n-2)^2}{16n^2} \frac{\phi^2}{|g|^2} \partial_\mu |g| \partial_\nu |g|  \Biggr]
- V \left( \frac{1}{2} \sqrt{\frac{n-2}{n-1}} |g|^{\frac{n-2}{4n}} \phi \right) \Biggr\}.
\label{WTDiff+Scalar Action}
\end{eqnarray}

From this action, the equations of motion for the scalar field and the metric tensor field
are respectively calculated to be
\begin{eqnarray}
&-& \frac{1}{8} \frac{n-2}{n-1} |g|^{\frac{1}{2}} \Biggl[ \frac{(n-2)(5n-2)}{8n^2} \frac{\phi}{|g|^2} (\partial_\rho |g|)^2
-  \frac{n-2}{2n} \frac{\phi}{|g|} \nabla_\rho \nabla^\rho |g| - 2 \nabla_\rho \nabla^\rho \phi  \Biggr]
\nonumber\\
&-& \frac{1}{2} \sqrt{\frac{n-2}{n-1}} |g|^{\frac{n-2}{4n}} V^\prime \left( \frac{1}{2} \sqrt{\frac{n-2}{n-1}} 
|g|^{\frac{n-2}{4n}} \phi \right) = 0,
\label{phi eq of WTDiff+Scalar}
\end{eqnarray}
with being $V^\prime (\phi) \equiv \frac{d V (\phi)}{d \phi}$, and
\begin{eqnarray}
R_{\mu\nu} - \frac{1}{n} g_{\mu\nu} R = T_{(g,\phi)\mu\nu} - \frac{1}{n} g_{\mu\nu} T_{(g,\phi)},
\label{Metric eq of WTDiff+Scalar}
\end{eqnarray}
where the energy-momentum tensor $T_{(g,\phi)\mu\nu}$ is defined as
\begin{eqnarray}
T_{(g,\phi)\mu\nu} &=& \frac{(n-2)(2n-1)}{4n^2} \frac{1}{|g|^2} \partial_\mu |g| \partial_\nu |g|
- \frac{n-2}{2n} \frac{1}{|g|} \nabla_\mu \nabla_\nu |g| 
\nonumber\\
&+& \frac{1}{4} \frac{n-2}{n-1} \left( \partial_\mu \phi + \frac{n-2}{4n} \frac{\phi}{|g|} \partial_\mu |g| \right) 
\left( \partial_\nu \phi + \frac{n-2}{4n} \frac{\phi}{|g|} \partial_\nu |g| \right).
\label{T_{(g,phi)}}
\end{eqnarray}
In deriving the energy-momentum tensor (\ref{T_{(g,phi)}}), we have used the equation of motion for $\phi$, 
(\ref{phi eq of WTDiff+Scalar}).

The energy-momentum tensor $T_{(g,\phi)\mu\nu}$ is not covariantly conserved, either,
but it is again possible to construct a covariantly conserved energy-momentum tensor as before. 
The covariantly conserved energy-momentum tensor is now given by
\begin{eqnarray}
T_{\mu\nu} &=& T_{(g,\phi)\mu\nu} + \frac{n-2}{2n} g_{\mu\nu} \left[ - \frac{5n-3}{4n} \frac{1}{|g|^2} 
(\partial_\rho |g|)^2 + \frac{1}{|g|} \nabla_\rho \nabla^\rho |g| \right]
\nonumber\\
&+& g_{\mu\nu} \left[ - \frac{1}{8} \frac{n-2}{n-1} \left( \partial_\rho \phi + \frac{n-2}{4n} \frac{\phi}{|g|} 
\partial_\rho |g| \right)^2 - |g|^{-\frac{1}{2}} V \left( \frac{1}{2} \sqrt{\frac{n-2}{n-1}} 
|g|^{\frac{n-2}{4n}} \phi \right) \right].
\label{new scalar T}
\end{eqnarray}
It turns out that this covariantly conserved energy-momentum tensor $T_{\mu\nu}$ satisfies the desired relation
\begin{eqnarray}
T_{\mu\nu} - \frac{1}{n} g_{\mu\nu} T = T_{(g,\phi)\mu\nu} - \frac{1}{n} g_{\mu\nu} T_{(g,\phi)}.
\label{Scalar-Critical relation of T}
\end{eqnarray}
Hence, although we add the scalar field to the WTDiff gravity, we can derive the standard Einstein equations 
(\ref{Einstein eq}) where the cosmological constant appears as an integration constant. 

To simplify the energy-momentum tensor, we will again select the unimodular condition (\ref{g=-1}) as a gauge condition 
for the Weyl symmetry. This choice of the gauge condition provides us with an enormous simplication since 
the energy-momentum tensor (\ref{new scalar T}) is reduced to the tractable form
\begin{eqnarray}
T_{\mu\nu} = \frac{1}{4} \frac{n-2}{n-1} \partial_\mu \phi \partial_\nu \phi
+ g_{\mu\nu} \left[ - \frac{1}{8} \frac{n-2}{n-1} (\partial_\rho \phi)^2 
- V \left( \frac{1}{2} \sqrt{\frac{n-2}{n-1}} \phi \right) \right].
\label{new scalar T in g=-1}
\end{eqnarray}

We are willing to go on to the study of cosmological solutions.
It is usually assumed that our universe is described in terms of an expanding, homogeneous and isotropic
Friedmann-Lemaitre-Robertson-Walker (FLRW) universe given by the line element 
\begin{eqnarray}
d s^2 &=& g_{\mu\nu} d x^\mu d x^\nu    \nonumber\\
&=& - d t^2 + a^2(t) \gamma_{ij}(x) d x^i d x^j,
\label{Line element 6-1}
\end{eqnarray}
where $a(t)$ is a scale factor and $\gamma_{ij}(x)$ is the spatial metric of the unit $(n-1)$-sphere,
unit $(n-1)$-hyperboloid or $(n-1)$-plane, and $i, j$ run over spatial coordinates ($i = 1, 2, \cdots, n-1$).
However, this metric ansatz does not satisfy the gauge condition (\ref{g=-1}) so the line element 
should be somewhat modified. A suitable modification, which satisfies the gauge condition (\ref{g=-1}),
is to consider the following line element;
\begin{eqnarray}
d s^2 = - N^2(t) d t^2 + a^2(t) (d x^i)^2,
\label{Line element 6-2}
\end{eqnarray}
where $N(t)$ is a lapse function and the spatial geometry is chosen to be the $(n-1)$-plane, i.e. the $(n-1)$-dimensional 
Euclidean space. Note that the existence of the lapse function $N(t)$ means that a time coordinate $t$ does not 
coincide with the proper time of particles at rest. With this line element, the gauge condition (\ref{g=-1}) provides 
a relation between the lapse function $N(t)$ and the scale factor  $a(t)$
\begin{eqnarray}
N(t) = a^{-(n-1)}(t).
\label{N vs a}
\end{eqnarray}

Given the line element (\ref{Line element 6-2}) and Eq. (\ref{N vs a}), it turns out 
that the non-vanishing components of the $\it{traceless}$ Einstein tensor defined as
$G^T_{\mu\nu} = R_{\mu\nu} - \frac{1}{n} g_{\mu\nu} R$ are given by
\begin{eqnarray}
G^T_{tt} &=& - \frac{(n-1)(n-2)}{n} \left[ \dot{H}  + (n-1) H^2 \right],  \nonumber\\
G^T_{ij} &=& - \frac{n-2}{n} a^{2n} \left[ \dot{H}  + (n-1) H^2 \right] \delta_{ij},
\label{traceless Eins-tensor 2}
\end{eqnarray}
where $H = \frac{\dot{a}}{a}$ is the Hubble parameter and we have defined $\dot{a} = \frac{d a(t)}{dt}$. 
In a similar way, the non-vanishing components of the $\it{traceless}$ energy-momentum tensor, which is defined as 
$T^T_{\mu\nu} = T_{\mu\nu} - \frac{1}{n} g_{\mu\nu} T$, read
\begin{eqnarray}
T^T_{tt} &=& \frac{n-2}{4n} (\dot{\phi})^2,  \nonumber\\
T^T_{ij} &=& \frac{1}{n-1} \frac{n-2}{4n} a^{2n} (\dot{\phi})^2 \delta_{ij},
\label{traceless stress-tensor}
\end{eqnarray}
where we have specified the scalar field $\phi$ to be spatially homogeneous, that is, $\phi = \phi(t)$.
As a result, the $\it{traceless}$ Einstein equations are cast to be a single equation
\begin{eqnarray}
\dot{H} + (n-1) H^2 = - \frac{1}{4(n-1)} (\dot{\phi})^2.
\label{Single eq}
\end{eqnarray}
Moreover, using the line element (\ref{Line element 6-2}) and Eq. (\ref{N vs a}), the equation of motion 
for the scalar matter field $\phi$, Eq. (\ref{phi eq of WTDiff+Scalar}), is simplified to be
\begin{eqnarray}
\ddot{\phi} + 2(n-1) H \dot{\phi} + 2 \sqrt{\frac{n-1}{n-2}} a^{-2(n-1)}
V^\prime \left( \frac{1}{2} \sqrt{\frac{n-2}{n-1}} \phi \right) = 0.
\label{Phi-eq 2}
\end{eqnarray}

It is of interest to see that the $\it{traceless}$ Einstein equations have yielded only
the single equation (\ref{Single eq}), which is similar to the Raychaudhuri equation or the first Friedmann 
equation \cite{Rubakov, Mukhanov}, which comes from all $ij$-components of the Einstein equations in general relativity 
though there is a slight difference in Eq. (\ref{Single eq}) which will be commented shortly. However, 
in the present formalism, the (second) Friedmann equation stemming from $00$-components of the Einstein equations is missing. 
In order to solve Eq. (\ref{Single eq}), we need the conservation law of the energy-momentum tensor.
In this respect, recall that in general relativity the first Friedmann equation can be viewed as a consequence
of the (second) Friedmann equation and covariant conservation of energy, so that the combination of 
the (second) Friedmann equation and the conservation law, supplemented by the equation of state
$p = p(\rho)$ (which will appear later), forms a complete system of equations that determines the
two unknown functions, the scale factor $a(t)$ and energy density $\rho$. In our formalism, instead of
the (second) Friedmann equation, we have to use the first Friedmann equation like Eq. (\ref{Single eq}).   

At this stage, we meet a new situation: As mentioned above, to solve the equation (\ref{Single eq}), 
we must set up the conservation law of the energy-momentum tensor as an additional equation. 
It is the covariantly conserved energy-momentum tensor $T_{\mu\nu}$ in Eq. (\ref{new scalar T in g=-1})
that we have to deal with in this context. So far the covariantly conserved energy-momentum tensors
are needed to make contact with the standard Einstein equations, but in the present situation, 
we must make use of the concrete expressions to find classical solutions.
Then, the non-vanishing components of $T_{\mu\nu}$ are easily evaluated to be 
\begin{eqnarray}
T^t \, _t &=& - \frac{1}{8} \frac{n-2}{n-1} a^{2(n-1)} (\dot{\phi})^2 
- V \left( \frac{1}{2} \sqrt{\frac{n-2}{n-1}} \phi \right) \equiv - \rho(t),
\nonumber\\
T^i \, _j &=& \left[ \frac{1}{8} \frac{n-2}{n-1} a^{2(n-1)} (\dot{\phi})^2 
- V \left( \frac{1}{2} \sqrt{\frac{n-2}{n-1}} \phi \right) \right] \delta^i \, _j
\equiv p(t) \delta^i \, _j,
\label{Comp-T}
\end{eqnarray}
where we have introduced energy density $\rho(t)$ and pressure $p(t)$ in a conventional way.
Using these expressions, the covariant conservation law $\nabla_\mu T^{\mu\nu} = 0$ leads to an equation
\begin{eqnarray}
\dot{\rho} + (n-1) H ( \rho + p ) = 0.
\label{rho-p}
\end{eqnarray}

To close the system of equations, which determines the dynamics of homogeneous and isotropic universe,
we have to specify the equation of state of matter as usual
\begin{eqnarray}
p = w \rho,
\label{w}
\end{eqnarray}
where $w$ is a certain constant. Of course, the equation of state is not a consequence of equations of 
our formalism, but should be determined by matter content in our universe. With the help of Eq. (\ref{w}),
Eq. (\ref{rho-p}) is exactly solved to be
\begin{eqnarray}
\rho(t) = \rho_0 a^{-(n-1)(w+1)}(t),
\label{rho-a}
\end{eqnarray}
where $\rho_0$ is an integration constant. Eqs. (\ref{rho-p})-(\ref{rho-a}) are the same expressions as
in general relativity.
Now, using Eqs. (\ref{Comp-T}), (\ref{w}) and (\ref{rho-a}), our Friedmann equation (\ref{Single eq})
is rewritten as
\begin{eqnarray}
\dot{H} + (n-1) H^2 = - \frac{w+1}{n-2} \rho_0 a^{-(n-1)(w+3)}.
\label{Friedmann eq}
\end{eqnarray}

Since it is difficult to find a general solution to this equation (\ref{Friedmann eq}), we will refer to only
special solutions which are physically interesting. Looking at the RHS in Eq. (\ref{Friedmann eq}), one
soon notices that at $w=-1$ and $w=-3$, specific situations occur. Actually, at $w=-1$, Eq. (\ref{Friedmann eq})
can be exactly integrated to be 
\begin{eqnarray}
a(t) = a_0 t^{\frac{1}{n-1}},
\label{a at w=-1}
\end{eqnarray}
where $a_0$ is an integration constant and this solution describes the decelerating universe in four dimensions
owing to $\ddot{a} <0$.

At the case $w=-3$, Eq. (\ref{Friedmann eq}) is reduced to the form
\begin{eqnarray}
\dot{H} + (n-1) H^2 = \frac{2}{n-2} \rho_0.
\label{Friedmann eq at w=-3}
\end{eqnarray}
This equation includes a special solution describing an exponentially expanding universe 
\begin{eqnarray}
a(t) = a_0 \e^{H_0 t},
\label{a at w=-3}
\end{eqnarray}
where $H_0$ is a constant defined as
\begin{eqnarray}
H_0 = \sqrt{\frac{2 \rho_0}{(n-1)(n-2)}}.
\label{H_0}
\end{eqnarray}

Finally, one can find a special solution such that the scale factor $a(t)$ is the form of polynomial 
in $t$ 
\begin{eqnarray}
a(t) = a_0 t^\alpha,
\label{Poly a}
\end{eqnarray}
where $\alpha$ is a constant to be determined by the Friedmann equation (\ref{Friedmann eq}). 
It is easy to verify that the constant $\alpha$ is given by
\begin{eqnarray}
\alpha = \frac{2}{(n-1)(w+3)},
\label{alpha}
\end{eqnarray}
so that in this case the scale factor takes the form
\begin{eqnarray}
a(t) = a_0 t^{\frac{2}{(n-1)(w+3)}}.
\label{Scale factor}
\end{eqnarray}
Then, the accelerating universe $\ddot{a}(t) > 0$ requires 
\begin{eqnarray}
w < \frac{-3n+5}{n-1},
\label{accelerating univ}
\end{eqnarray}
while the decelerating universe does
\begin{eqnarray}
w > \frac{-3n+5}{n-1}.
\label{decelerating univ}
\end{eqnarray}

One might wonder how the obtained solutions are related to solutions in general relativity.
In particular, in general relativity we are familiar with the fact that the case $w=-1$ corresponds 
to the cosmological constant and the solution is then an exponentially expanding universe whereas
in our case the corresponding solution belongs to the case $w=-3$, which appears to be strange. But this is 
just an illusion since we do not use the conventional form (\ref{Line element 6-1}) of the line element 
but the line element (\ref{Line element 6-2}) involving the nontrivial lapse function $N(t)$.

In order to show that our result coincides with that in general relativity, let us focus our
attention to the Friedmann equation (\ref{Single eq}). By means of Eq. (\ref{Comp-T}), this equation
is rewritten as    
\begin{eqnarray}
\dot{H} + (n-1) H^2 = - \frac{1}{n-2} N^2 (\rho + p),
\label{Single eq 2}
\end{eqnarray}
where we recovered the lapse function $N(t)$ by using Eq. (\ref{N vs a}).
  
On the other hand, with the conventional notation of the energy-momentum tensor
\begin{eqnarray}
T^\mu \, _\nu = diag ( -\rho, p, \cdots, p ),
\label{Cov-T}
\end{eqnarray}
and the line element (\ref{Line element 6-2}), the Einstein equations in general relativity 
\begin{eqnarray}
G^\mu \, _\nu \equiv R^\mu \, _\nu - \frac{1}{2} \delta^\mu \, _\nu R = T^\mu \, _\nu,
\label{Cov-Ein-eq}
\end{eqnarray}
become a set of the Friedmann equations   
\begin{eqnarray}
&{}& H^2 = \frac{2}{(n-1)(n-2)} N^2 \rho,  
\label{Cov-Fried-eq 1} \\
&{}& \dot{H} + \frac{n-1}{2} H^2 - \frac{\dot{N}}{N} H = - \frac{1}{n-2} N^2 p.
\label{Cov-Fried-eq 2}
\end{eqnarray}
By using Eq. (\ref{N vs a}), Eq. (\ref{Cov-Fried-eq 2}) is written as 
\begin{eqnarray}
\dot{H} + \frac{3(n-1)}{2} H^2 = - \frac{1}{n-2} N^2 p.
\label{Cov-Fried-eq 2-1}
\end{eqnarray}
Eq. (\ref{Cov-Fried-eq 1}) allows us to rewrite this equation to the form
\begin{eqnarray}
\dot{H} + (n-1) H^2 = - \frac{1}{n-2} N^2 (\rho + p),
\label{Cov-Fried-eq 2-2}
\end{eqnarray}
which precisely coincides with our Friedmann equation (\ref{Single eq 2}). This demonstration
clearly indicates that our cosmological solution is just equivalent to that of
general relativity specified in such a way that the line element is (\ref{Line element 6-2})
and the lapse function is given by Eq. (\ref{N vs a}).

\section{Discussions}

In this article, we have clarified various classical aspects of the Weyl transverse (WTDiff) gravity
in a general space-time dimension. We have found that Schwarzschild black hole is a classical solution
to the equations of motion of the WTDiff gravity when expressed in the Cartesian coordinate system.
We have also shown that the Reissner-Nordstrom black hole is a solution in the same coordinate system
in four space-time dimensions. The generalization to higher space-time dimensions has required us to 
extend the conventional Maxwell action to the Weyl-invariant action since the Maxwell action is
invariant under the Weyl (local conformal) transformation only in four dimensions. It is of interest
that even in such an extended electro-magnetic action plus the WTDiff gravity action in higher dimensions 
there is a charged black hole solution which shares the whole features with the conventional Reissner-Nordstrom 
charge black hole solution in four dimensions. Furthermore, we have investigated the 
Friedmann-Lemaitre-Robertson-Walker (FLRW) cosmology and seen that the FLRW cosmology is a classical solution
when the shift factor has a nontrivial scale factor and the spacial geometry is flat.    

In the classical analysis of the WTDiff gravity, a novel feature is the classical relation among three gravitational 
theories, those are, the conformally invariant scalar-tensor gravity, Einstein's general relativity and
the Weyl transverse (WTDiff) gravity, in a general space-time dimension. To put it concretely, starting with 
the conformally invariant scalar-tensor gravity which is invariant under both the local Weyl transformation 
and diffeomorphisms (Diff), we have gauge-fixed the longitudinal diffeomorphism, by which the full diffeomorphisms 
(Diff) are broken to the transverse diffeomorphisms (TDiff), and we have obtained the WTDiff gravity. It is explicitly 
verified that not only the resultant action of the WTDiff gravity but also its equations of motion are invariant 
under both the local Weyl transformation and TDiff. On the other hand, beginning with the conformally invariant 
scalar-tensor gravity and gauge-fixing the Weyl transformation has yielded general relativity which is invariant 
under Diff. In this sense, the three gravitational theories are classically equivalent and we then conjecture that 
this equivalence holds even in the quantum regime. In other words, the conformally invariant scalar-tensor gravity
is the underlying theory with the maximum symmetry behind Einstein's general relativity and the WTDiff gravity.

As a bonus, the equivalence among the three theories has made it possible to construct covariantly conserved energy-momentum
tensors, by which we can prove that the traceless Einstein equations in the WTDiff gravity become equivalent to
the standard Einstein equations in general relativity. Here one of the most remakable things is that the cosmological
constant emerges as an integration constant. This interesting phenomenon has been already observed in unimodular gravity
and expected to lead to a resolution to the cosmological constant problem. However, afterwards, it was revealed that
this is not indeed the case by the following reason: In unimodular gravity, the unimodular condition
plays an important role and this condition must be properly implemented by the method of Lagrange multiplier. Then, it turns
out that the Lagrange multiplier field is nothing but the cosmological constant and receives huge radiative corrections.

On the other hand, in the WTDiff gravity under consideration, we have a chance of utilizing the phenomenon of the emergence
of the cosmological constant as an integration constant for solving the cosmological constant problem. In the WTDiff gravity,
we have neither additional conditions like the unimodular condition nor Lagrange multiplier fields, so we have no counterpart
of the cosmological constant in the action. Moreover, the Weyl symmetry forbids the emergence of the cosmological
constant of dimension zero in a quantum effective action, and if it were not violated at the quantum level, the cosmological 
constant appearing as an integration constant in the Einstein equations would keep its classical value in all energy scales. 
In this sense, the cosmological constant in the WTDiff gravity is radiatively stable. Thus, important remaining works amount to 
giving a proof that the {\it fake} Weyl symmetry is not broken by quantum effects and determining the initial value of the 
cosmological constant from some still unknown principle.    

In a pioneering paper by Englert et al. \cite{Englert}, it is stated that the Weyl symmetry in the conformally invariant 
scalar-tensor gravity is free of Weyl anomaly when the Weyl symmetry is spontaneously broken, and this situation is unchanged 
when the Weyl invariant matter fields are incorporated into the theory. Here "spontaneously broken" needs an explanation. 
Usually, it is necessary to have a Higgs potential to trigger the spontaneous symmetry breakdown, but it is in general difficult to 
set up such a potential for breaking the Weyl symmetry. Thus, as commented on around the end of Section 3, the meaning of 
"spontaneously broken" should be understood in the sense that the spurion field $\varphi$ is assumed to be divided into two terms, 
$\varphi = \langle \varphi \rangle + \sigma$ where $\langle \varphi \rangle$ is the vacuum expectation value and $\sigma$ is 
the Goldstone boson restoring conformal symmetry, respectively. Then, the key technical idea in \cite{Englert}-\cite{Ghilencea2}
is that the vacuum expectation value $\langle \varphi \rangle$ plays a role as the renormalization scale instead of the conventional 
fixed renormalization scale, by which the Weyl invariant effective action can be obtained. Our conjecture that the fake Weyl symmetry 
has no anomaly is interpreted as a supplementary statement from the symmetry side, which supports this technical idea.
      
Anyway, as an important feature problem, we must understand quantum aspects of the WTDiff gravity. This is a very important step 
for the cosmological constant problem.  We wish to consider this problem in near future.

\begin{flushleft}
{\bf Acknowledgements}
\end{flushleft}
This article is delicated to one of my friends, Mario Tonin who suddenly passed away this April. 
We have collaborated with pure spinor formalism of superstring theories and completed several papers.
I thank him very much for continuous encouragements and warm hospitality in Padova. 
We are also grateful to the generous hospitality of the Galileo Galilei Institute of Padova University 
where part of this work has been done.
This work is supported in part by the Grant-in-Aid for Scientific 
Research (C) No. 16K05327 from the Japan Ministry of Education, Culture, 
Sports, Science and Technology.

\newpage

\begin{flushleft}
{\bf Appendices}
\end{flushleft}
\appendix

\renewcommand{\theequation}{A.\arabic{equation}}
\setcounter{equation}{0}

\section{Notation and conventions}

\subsection{Gravity}

We follow notation and conventions by Misner et al.'s textbook \cite{MTW}, for instance, the flat Minkowski 
metric $\eta_{ab} = diag(- 1, 1, 1, 1)$, the Riemann curvature tensor $R^\mu \, _{\nu\alpha\beta} 
= \partial_\alpha \Gamma^\mu_{\nu\beta} - \partial_\beta \Gamma^\mu_{\nu\alpha} 
+ \Gamma^\mu_{\sigma\alpha} \Gamma^\sigma_{\nu\beta} - \Gamma^\mu_{\sigma\beta} \Gamma^\sigma_{\nu\alpha}$, 
and the Ricci tensor $R_{\mu\nu} = R^\alpha \, _{\mu\alpha\nu}$. The Latin indices label the flat space-time
coordinates while the Greek ones run over the curved space-time coordinates.
The reduced Planck mass is defined as $M_p = \sqrt{\frac{c \hbar}{8 \pi G}} 
= 2.4 \times 10^{18} GeV$. Throughout this article, we adopt the reduced Planck units where we set 
$c = \hbar = M_p = 1$. In this units, all quantities become dimensionless. 
Finally, note that in the reduced Planck units, the Einstein-Hilbert Lagrangian density takes the form
${\cal L}_{EH} = \frac{1}{2} \sqrt{-g} R$.

\subsection{Spinor}

In this subsection, we gather some notation and definitions relevant to spinor fields. The Dirac spinor
$\psi$ is a $2^{[\frac{n}{2}]}$ dimensional spinor where $[\frac{n}{2}]$ is the Gauss symbol.  The Clifford
algebra is defined as $\{ \gamma^a, \gamma^b \} = 2 \eta^{ab}$. The gamma matrices in a curved space-time
are related to those in a flat space-time with the help of the vielbein by $\gamma^\mu = e_a ^\mu \gamma^a$.
The metric tensor $g_{\mu\nu}$ is composed of the vielbein $e_a ^\mu$ by the conventional relation
$g_{\mu\nu} = \eta^{ab} e_{a \mu} e_{b \nu}$.  Therefore, we have $\{ \gamma^\mu, \gamma^\nu \} = 2 g^{\mu\nu}$.

In writing down the Dirac action, we need to define the Dirac adjoint and the covariant derivative. The Dirac
adjoint is defined as $\bar \psi = - i \psi^\dagger \gamma_{a=0} = i \psi^\dagger \gamma^{a=0}$ where $\gamma_{a=0}$
or $\gamma^{a=0}$ denotes the zero component of the flat space-time gamma matrices. Using the spin connection
$\omega_\mu ^{ab}$, the covariant derivative is of form  
\begin{eqnarray}
D_\mu \psi = ( \partial_\mu + \frac{1}{4} \omega_\mu ^{ab} \gamma_{ab} ) \psi,
\label{Cov-der}
\end{eqnarray}
where $\gamma_{ab} = \frac{1}{2} [ \gamma_a, \gamma_b ]$. Similarly, the covariant derivative for the Dirac adjoint
can be derived from (\ref{Cov-der}) to be 
\begin{eqnarray}
\bar \psi \overleftarrow D_\mu = \bar \psi ( \overleftarrow \partial_\mu - \frac{1}{4} \omega_\mu ^{ab} \gamma_{ab} ).
\label{Adj-cov-der}
\end{eqnarray}

We will use the torsion-free spin connection. Then, it is defined through the Ricci rotation coefficient as
\begin{eqnarray}
\omega_{a, bc} = e_a ^\mu \omega_{\mu, bc} 
= \frac{1}{2} ( \Delta_{a, bc} - \Delta_{b, ca} + \Delta_{c, ba}  ) 
= - \omega_{a, cb},
\label{Spin connection}
\end{eqnarray}
where the Ricci rotation coefficient is defined as 
\begin{eqnarray}
\Delta_{a, bc} = - \Delta_{a, cb} 
= ( e_b ^\mu e_c ^\nu - e_c ^\mu e_b ^\nu ) \partial_\nu e_{a \mu} 
= - e_{a \mu} ( e_c ^\nu \partial_\nu e_b ^\mu - e_b ^\nu \partial_\nu e_c ^\mu ).
\label{Ricci-rot}
\end{eqnarray}

\section{Proof of invariance}

\renewcommand{\theequation}{B.\arabic{equation}}
\setcounter{equation}{0}

In this appendix, let us explicitly show that the action (\ref{WTDiff Action}) and the equations of motion 
(\ref{Eq from WTDiff Action}) are invariant under the Weyl transformation (\ref{Weyl transf}) and 
the transverse group of diffeomorphisms.

For this purpose, let us explain the transverse diffeomorphisms (TDiff) in more detail. Under the general 
coordinate transformation or Diff, the metric tensor transforms as
\begin{eqnarray}
g_{\mu\nu}(x) \rightarrow g_{\mu\nu}^\prime(x^\prime) = \frac{\partial x^\alpha}{\partial x^{\mu \prime}}
\frac{\partial x^\beta}{\partial x^{\nu \prime}} g_{\alpha\beta}(x) \equiv J^\alpha_{\mu \prime}
J^\beta_{\nu \prime} g_{\alpha\beta}(x),
\label{Metric diff}
\end{eqnarray}
where the Jacobian matrix $J^\alpha_{\mu \prime}$, which is defined as $J^\alpha_{\mu \prime} = 
\frac{\partial x^\alpha}{\partial x^{\mu \prime}}$, was introduced. Denoting the determinant of the Jacobian matrix as
$J = \det J^\alpha_{\mu \prime} = \det \frac{\partial x^\alpha}{\partial x^{\mu \prime}}$, taking the
determinant of Eq. (\ref{Metric diff}) produces 
\begin{eqnarray}
g^\prime(x^\prime) = J^2(x) g(x).
\label{J}
\end{eqnarray}
Then, the transverse diffeomorphisms (TDiff), or equivalently the volume preserving diffeomorphisms,
are defined as a subgroup of the full diffeomorphisms such that the determinant of the Jacobian matrix is the unity
\begin{eqnarray}
J(x) = 1.
\label{Unimodular J}
\end{eqnarray}
With this condition (\ref{Unimodular J}), the volume element is preserved under Diff, and Eq. (\ref{J}) shows
that $g(x)$ is a dimensionless scalar field.  In the infinitesimal form of diffeomorphisms $x^\mu \rightarrow x^{\mu \prime} 
= x^\mu - \xi^\mu(x)$, using Eq. (\ref{Unimodular J}), TDiff can be expressed in terms of Eq. (\ref{TDiff}) 
since we can derive the following equation 
\begin{eqnarray}
1 = J(x) = \mathrm{det} \frac{\partial x^\alpha}{\partial x^{\mu \prime}}
= \mathrm{det} \left( \delta_\mu^\alpha + \partial_\mu \xi^\alpha \right)
= \mathrm{e}^{\mathrm Tr \log \left( \delta_\mu^\alpha + \partial_\mu \xi^\alpha \right)}
= \mathrm{e}^{\partial_\mu \xi^\mu}.
\label{Logic of TDiff}
\end{eqnarray}

Armed with the knowledge of TDiff, we are ready to show explicitly that the action (\ref{WTDiff Action}) and 
the equations of motion (\ref{Eq from WTDiff Action}) of the WTDiff gravity are indeed invariant under both TDiff
and Weyl transformation. In fact, under Diff, the Lagrangian density of (\ref{WTDiff Action}) is transformed as
\begin{eqnarray}
{\cal L}^\prime(x^\prime) 
= \frac{1}{2} |J^2 g|^{\frac{1}{n}} \left[ R + \frac{(n-1)(n-2)}{4n^2} \frac{1}{|g|^2}
g^{\mu\nu} (\partial_\mu |g| + \frac{2 |g|}{J} \partial_\mu J) 
(\partial_\nu |g| + \frac{2 |g|}{J} \partial_\nu J)  \right].
\label{Diff of WTDiff Action}
\end{eqnarray}
It is obvious that the Lagrangian density ${\cal L}$ is not invariant under Diff owing to
the presence of the terms with $J$ while it is invariant under TDiff because of Eq. (\ref{Unimodular J}), 
which means that TDiff are in fact a symmetry of the action (\ref{WTDiff Action}) of the WTDiff gravity. 
Now let us show that the traceless Einstein equations (\ref{Eq from WTDiff Action}) are also
invariant under TDiff. To do so, let us perform the general coordinate transformation to 
Eq. (\ref{Eq from WTDiff Action}) whose result is described as 
\begin{eqnarray}
G_{\mu\nu}^{T \prime} - T_{(g) \mu\nu}^{T \prime} 
&=& J_{\mu \prime}^\alpha J_{\nu \prime}^\beta \Biggl\{  G_{\alpha\beta}^T - T_{(g) \alpha\beta}^T 
+ \frac{n-2}{2n} \biggl[ \frac{1}{n} \frac{1}{J|g|} (\partial_\alpha J \partial_\beta |g|
+ \partial_\beta J \partial_\alpha |g|)  
\nonumber\\
&+& \frac{2(1-n)}{n} \frac{1}{J^2} \partial_\alpha J \partial_\beta J 
+ \frac{2}{J} D_\alpha D_\beta J \biggr] 
- \frac{n-2}{n^2} \biggl[ \frac{1}{n} \frac{1}{J|g|} \partial_\rho J \partial^\rho |g|
\nonumber\\
&+& \frac{1-n}{n} \frac{1}{J^2} (\partial_\rho J)^2 + \frac{1}{J} D_\rho D^\rho J \biggr] 
g_{\alpha\beta} \Biggr\}.
\label{Diff of Eq of motion}
\end{eqnarray}
From this expression, we see that (\ref{Eq from WTDiff Action}) is not invariant under Diff, but with $J=1$, 
that is, under TDiff, it becomes invariant. 

Next, we will prove the Weyl invariance of the action (\ref{WTDiff Action}) and the equations of motion 
(\ref{Eq from WTDiff Action}). Under the Weyl transformation (\ref{Weyl transf}), the Lagrangian density 
of (\ref{WTDiff Action}) is changed as
\begin{eqnarray}
{\cal L}^\prime = {\cal L} - (n-1) \partial_\mu \left( |g|^{\frac{1}{n}} g^{\mu\nu} \frac{1}{\Omega} 
\partial_\nu \Omega \right),
\label{Weyl transf of L}
\end{eqnarray}
which implies that the WTDiff gravity is invariant under the Weyl transformation up to a surface term.     
Now, under the Weyl transformation, the traceless Einstein tensor $G_{\mu\nu}^T$
and $T_{(g) \mu\nu}^T$ are transformed by the same quantity
\begin{eqnarray}
G_{\mu\nu}^{T \prime} &=& G_{\mu\nu}^T + A_{\mu\nu}^T,   \nonumber\\
T_{(g) \mu\nu}^{T \prime} &=& T_{(g) \mu\nu}^T + A_{\mu\nu}^T,
\label{Weyl transf of Eq}
\end{eqnarray}
where $A_{\mu\nu}^T$ is defined as
\begin{eqnarray}
A_{\mu\nu}^T = 2(n-2) \frac{1}{\Omega^2} \left[ \partial_\mu \Omega \partial_\nu \Omega 
- \frac{1}{n} g_{\mu\nu} (\partial_\rho \Omega)^2 \right]
-(n-2) \frac{1}{\Omega} \left[ \nabla_\mu \nabla_\nu \Omega 
- \frac{1}{n} g_{\mu\nu} \nabla_\rho \nabla^\rho \Omega \right]. 
\label{A}
\end{eqnarray}
It is therefore obvious that Eq. (\ref{Eq from WTDiff Action}) is invariant under the Weyl transformation.

\section{Derivations of Eq. (\ref{Eq from WTDiff Action})}

\renewcommand{\theequation}{C.\arabic{equation}}
\setcounter{equation}{0}

In this appendix, we will present two different derivations of the equations of motion 
(\ref{Eq from WTDiff Action}) for the metric tensor in the WTDiff gravity.

\subsection{Derivation from Eq. (\ref{g-Eq of motion of conf-ST 2})}

This derivation method utilizes the equivalence relation between the conformally invariant
scalar-tensor gravity and the WTDiff gravity via the gauge fixing procedure, and the
fact that the equations of motion for the metric tensor in the WTDiff gravity are traceless
equations.

As mentioned in the article, the equations of motion in the WTDiff gravity is entirely described 
in Eq. (\ref{g-Eq of motion of conf-ST 2}), or equivalently Eq. (\ref{g-Eq of motion of conf-ST}).
The equivalence between the conformally invariant scalar-tensor gravity and the WTDiff gravity 
via the gauge fixing procedure demands that the equations of motion in the WTDiff gravity should
be obtained from Eq. (\ref{g-Eq of motion of conf-ST 2}) by substituting the gauge condition (\ref{WTDiff gauge}).
After a straightforward calculation, we find that 
\begin{eqnarray}
G_{\mu\nu} &=& \frac{(n-2)(2n-1)}{4 n^2} \frac{1}{|g|^2} \partial_\mu |g| \partial_\nu |g| 
- \frac{n-2}{2n} \frac{1}{|g|} \nabla_\mu \nabla_\nu |g|
\nonumber\\
&-& \frac{(n-2)(5n-3)}{8 n^2} g_{\mu\nu} \frac{1}{|g|^2} \left( \partial_\rho |g| \right)^2
+ \frac{n-2}{2n} g_{\mu\nu} \frac{1}{|g|} \nabla_\rho \nabla^\rho |g|.
\label{Appendix B1}
\end{eqnarray}
It is easy to see that taking its traceless part, i.e. calculating $G_{\mu\nu}^T \equiv R_{\mu\nu}
- \frac{1}{n} g_{\mu\nu} R$, yields the equations of motion in the WTDiff gravity,
Eq. (\ref{Eq from WTDiff Action}) with the definition of the energy-momentum tensor (\ref{T(g)}).

\subsection{Derivation from variation of WTDiff gravity action (\ref{WTDiff Action})}

In this subsection, we will derive the equations of motion (\ref{Eq from WTDiff Action}) of the WTDiff 
gravity by taking the variation for the metric tensor step by step.

Let us first divide the action of the WTDiff gravity, Eq. (\ref{WTDiff Action}) into two parts
\begin{eqnarray}
S = S_R + S_g = \int d^n x {\cal{L}}_R +  \int d^n x {\cal{L}}_g,
\label{Append-WTDiff Action}
\end{eqnarray}
where we have defined 
\begin{eqnarray}
{\cal{L}}_R = \frac{1}{2} |g|^{\frac{1}{n}} R, \quad
{\cal{L}}_g = \frac{(n-1)(n-2)}{8n^2} |g|^{\frac{1}{n} - 2}
g^{\mu\nu} \partial_\mu |g| \partial_\nu |g|.
\label{Append-WTDiff Lagr}
\end{eqnarray}
Using the formulae 
\begin{eqnarray}
\delta |g| = - |g| g_{\mu\nu} \delta g^{\mu\nu}, \quad
\delta R = R_{\mu\nu} \delta g^{\mu\nu} + ( g_{\mu\nu} \Box - \nabla_\mu \nabla_\nu ) 
\delta g^{\mu\nu}.
\label{Formulae}
\end{eqnarray}
the metric variation of ${\cal{L}}_R$ reads
\begin{eqnarray}
\delta {\cal{L}}_R = \frac{1}{2} |g|^{\frac{1}{n}}  G_{\mu\nu}^T \delta g^{\mu\nu}
+ \frac{1}{2} |g|^{\frac{1}{n}} ( g_{\mu\nu} \Box - \nabla_\mu \nabla_\nu ) \delta g^{\mu\nu}.
\label{Var-L_R}
\end{eqnarray}

Next, let us divide the second term in (\ref{Var-L_R}) into two parts and evaluate each term
separately
\begin{eqnarray}
I_1 = \frac{1}{2} |g|^{\frac{1}{n}} g_{\mu\nu} \Box \delta g^{\mu\nu}, \quad
I_2 = - \frac{1}{2} |g|^{\frac{1}{n}} \nabla_\mu \nabla_\nu \delta g^{\mu\nu}.
\label{I}
\end{eqnarray}
In what follows, to convert the covariant deivative $\nabla_\mu$ to the partial derivative $\partial_\mu$ 
we repeatedly use the well-known formula 
\begin{eqnarray}
|g|^{\frac{1}{2}} \nabla_\mu A^\mu = \partial_\mu \left( |g|^{\frac{1}{2}} A^\mu \right),
\label{Useful formula}
\end{eqnarray}
where $A^\mu$ is a generic vector field which includes $\partial^\mu |g|$ and $\nabla_\nu \delta g^{\mu\nu}$
etc. We will give a detailed derivation of $I_1$ below and only give the result of $I_2$ since the calculation 
of $I_2$ is similar to that of $I_1$. Neglecting total derivative terms and using the formula (\ref{Useful formula})
twice, we can proceed to calculate $I_1$ as follows:
\begin{eqnarray}
I_1 &=& \frac{1}{2} |g|^{\frac{1}{n} - \frac{1}{2}} |g|^{\frac{1}{2}} \nabla_\rho \left( g_{\mu\nu} g^{\rho\sigma}
\nabla_\sigma \delta g^{\mu\nu} \right)      \nonumber\\
&=& \frac{1}{2} |g|^{\frac{1}{n} - \frac{1}{2}} \partial_\rho \left( |g|^{\frac{1}{2}} g_{\mu\nu} g^{\rho\sigma}
\nabla_\sigma \delta g^{\mu\nu} \right)      \nonumber\\
&=& - \frac{1}{2} \left( \frac{1}{n} - \frac{1}{2} \right) |g|^{\frac{1}{n} - \frac{3}{2}} \partial_\rho
|g| |g|^{\frac{1}{2}} g_{\mu\nu} g^{\rho\sigma} \nabla_\sigma \delta g^{\mu\nu}   \nonumber\\
&=& \frac{n-2}{4n} |g|^{\frac{1}{n} - \frac{3}{2}} \left[ |g|^{\frac{1}{2}} \nabla_\sigma \left( \partial_\rho |g|  
g_{\mu\nu} g^{\rho\sigma} \delta g^{\mu\nu} \right) 
-  |g|^{\frac{1}{2}} \nabla_\rho \nabla_\sigma |g| g_{\mu\nu} g^{\rho\sigma} \delta g^{\mu\nu} \right]  \nonumber\\
&=& \frac{n-2}{4n} |g|^{\frac{1}{n} - \frac{3}{2}} \left[ \partial_\sigma \left( |g|^{\frac{1}{2}} \partial_\rho |g|  
g_{\mu\nu} g^{\rho\sigma} \delta g^{\mu\nu} \right) 
- |g|^{\frac{1}{2}} \nabla_\rho \nabla_\sigma |g| g_{\mu\nu} g^{\rho\sigma} \delta g^{\mu\nu} \right]  \nonumber\\
&=& \frac{n-2}{4n} \left[ - \left( \frac{1}{n} - \frac{3}{2} \right) |g|^{\frac{1}{n} - \frac{5}{2}}  \partial_\sigma |g| 
|g|^{\frac{1}{2}} \partial_\rho |g| g_{\mu\nu} g^{\rho\sigma} \delta g^{\mu\nu}  
- |g|^{\frac{1}{n} - 1} g_{\mu\nu} \nabla_\rho \nabla^\rho |g| \delta g^{\mu\nu} \right]  \nonumber\\
&=& \frac{n-2}{4n} |g|^{\frac{1}{n}} g_{\mu\nu} \delta g^{\mu\nu} \left[  \frac{3n-2}{2n} \frac{1}{|g|^2} 
(\partial_\rho |g|)^2  - \frac{1}{|g|} \nabla_\rho \nabla^\rho |g| \right].
\label{I_1}
\end{eqnarray}
In a perfectly similar way, we have
\begin{eqnarray}
I_2 = - \frac{n-2}{4n} |g|^{\frac{1}{n}} \delta g^{\mu\nu} \left[  \frac{3n-2}{2n} \frac{1}{|g|^2} 
\partial_\mu |g| \partial_\nu |g| - \frac{1}{|g|} \nabla_\mu \nabla_\nu |g| \right].
\label{I_2}
\end{eqnarray}
Then from Eqs. (\ref{Var-L_R}), (\ref{I_1}) and (\ref{I_2}), the variation of $S_R$ with respect to
the metric tensor becomes  
\begin{eqnarray}
\frac{\delta S_R}{\delta g^{\mu\nu}} &=& \frac{1}{2} |g|^{\frac{1}{n}}  G_{\mu\nu}^T 
+ \frac{n-2}{4n} |g|^{\frac{1}{n}} g_{\mu\nu} \left[  \frac{3n-2}{2n} \frac{1}{|g|^2} 
(\partial_\rho |g|)^2  - \frac{1}{|g|} \nabla_\rho \nabla^\rho |g| \right]
\nonumber\\
&-& \frac{n-2}{4n} |g|^{\frac{1}{n}} \left[  \frac{3n-2}{2n} \frac{1}{|g|^2} 
\partial_\mu |g| \partial_\nu |g| - \frac{1}{|g|} \nabla_\mu \nabla_\nu |g| \right].
\label{Var-S_R}
\end{eqnarray}

The variation of $S_g$ with respect to the metric tensor can be calculated in a similar manner to be
\begin{eqnarray}
\frac{\delta S_g}{\delta g^{\mu\nu}}   
&=& \frac{(n-1)(n-2)}{8n^2} |g|^{\frac{1}{n}} \Biggl\{ \frac{1}{|g|^2} \partial_\mu |g| \partial_\nu |g|
+ g_{\mu\nu} \Biggl[ - \frac{3n-1}{n} \frac{1}{|g|^2} (\partial_\rho |g|)^2 
\nonumber\\
&+& \frac{2}{|g|} \nabla_\rho \nabla_\rho |g| \Biggr] \Biggr\}.
\label{Var-g_R}
\end{eqnarray}
It is easy to check that adding the two results (\ref{Var-S_R}) and (\ref{Var-g_R}) leads to the equations
of motion of the WTDiff gravity
\begin{eqnarray}
\frac{\delta S}{\delta g^{\mu\nu}} = \frac{\delta S_R}{\delta g^{\mu\nu}} + \frac{\delta S_g}{\delta g^{\mu\nu}}
= \frac{1}{2} |g|^{\frac{1}{n}}  \left( G_{\mu\nu}^T - T_{\mu\nu}^T \right).
\label{Var-S}
\end{eqnarray}

\section{Covariantly conserved energy-momentum tensors}

\renewcommand{\theequation}{D.\arabic{equation}}
\setcounter{equation}{0}

In this article, we mainly work with the Weyl transverse (WTDiff) gravity which is not invariant 
under the general coordinate transformation (Diff) but only invariant under the Weyl transformation 
and TDiff. We find that the energy-momentum tensor derived from the WTDiff gravity
is not covariantly conserved, thereby making it unclear to make a connection with the standard Einstein
equations. However, as shown in this paper, the WTDiff gravity can be obtained by gauge-fixing
the longitudinal diffeomorphism existing in the conformally invariant scalar-tensor gravity,
which is generally covariant, so there should be a covariantly conserved energy-momentum tensor.
In this appendix, for completeness, we will give a (well-known) proof for the existence 
of the covariantly conserved energy-momentum tensor if the underlying gravitational theory is 
invariant under the general coordinate transformation (Diff). 

Suppose that a generic action $S$ is invariant under Diff 
\begin{eqnarray}
S = \int d^n x \sqrt{-g} \cal{L}.
\label{S}
\end{eqnarray}
Under Diff, the metric tensor transforms as
\begin{eqnarray}
\delta g^{\mu\nu} = \nabla^\mu \xi^\nu + \nabla^\nu \xi^\mu, 
\label{Metric GCT}
\end{eqnarray}
where $\xi^\mu$ is a local parameter of Diff. Under Diff, the action $S$ is transformed into
\begin{eqnarray}
\delta S = - \int d^n x \sqrt{-g} T_{\mu\nu} \nabla^\mu \xi^\nu, 
\label{delta S}
\end{eqnarray}
where the energy-momentum tensor $T_{\mu\nu}$ is defined as
\begin{eqnarray}
T_{\mu\nu} &=& - \frac{2}{\sqrt{-g}} \frac{\delta (\sqrt{-g} \cal{L})}{\delta g^{\mu\nu}}
\nonumber\\
&=& -2 \frac{\delta \cal{L}}{\delta g^{\mu\nu}} + g_{\mu\nu} \cal{L}. 
\label{def T}
\end{eqnarray}
By using the formula (\ref{Useful formula}) and integrating by parts, Eq. (\ref{delta S}) can be recast 
to the form
\begin{eqnarray}
\delta S = \int d^n x \sqrt{-g} \nabla_\mu T^{\mu\nu} \xi_\nu, 
\label{delta S 2}
\end{eqnarray}
from which we can arrive at the covariant conservation law of the the energy-momentum tensor
\begin{eqnarray}
\nabla_\mu T^{\mu\nu} = 0. 
\label{Conserv-law}
\end{eqnarray}
Let us note that only the general coordinate invariance of the action plays a critical role 
in this proof.


\end{document}